\documentclass[final]{siamart1116}
\usepackage{amssymb}
\usepackage{eurosym}
\usepackage{lipsum}
\usepackage{amsfonts}
\usepackage{graphicx}
\usepackage{epstopdf}
\usepackage{algorithmic}
\usepackage{amsopn}

\ifpdf
  \DeclareGraphicsExtensions{.eps,.pdf,.png,.jpg}
\else
  \DeclareGraphicsExtensions{.eps}
\fi
\numberwithin{theorem}{section}
\newcommand{\TheTitle}{Convexification of a Hyperbolic Inverse
Problem}
\newcommand{\TheAuthors}{M.V.Klibanov J.Li and W.Zhang}

\headers{\TheTitle}{\TheAuthors}

\ifpdf
\hypersetup{
  pdftitle={\TheTitle},
  pdfauthor={\TheAuthors}
}
\fi

\begin{document}

\title{Convexification for the Inversion of a Time Dependent Wave Front in a
Heterogeneous Medium\thanks{%
Submitted to the editors DATE. 
\funding{The work of Klibanov was supported by US Army Research Laboratory and US Army Research Office grant W911NF-19-1-0044.
The work of Li and Zhang was partially supported by the NSF of China under the grant No. 11571161 and 11731006, the Shenzhen Sci-Tech Fund No.
JCYJ20160530184212170 and JCYJ20170818153840322.}}}
\author{ Michael V. Klibanov\thanks{%
Department of Mathematics and Statistics, University of North Carolina at
Charlotte, Charlotte, NC 28223, USA (mklibanv@uncc.edu)} \and Jingzhi Li\thanks{%
Corresponding author. Department of Mathematics, Southern University of
Science and Technology (SUSTech), 1088 Xueyuan Boulevard, University Town of
Shenzhen, Xili, Nanshan, Shenzhen, Guangdong Province, P.R.China
(li.jz@sustc.edu.cn)} \and Wenlong Zhang\thanks{%
Department of Mathematics, Southern University of Science and Technology
(SUSTech), 1088 Xueyuan Boulevard, University Town of Shenzhen, Xili,
Nanshan, Shenzhen, Guangdong Province, P.R.China (zhangwl@sustc.edu.cn)} }
\date{}
\maketitle

\begin{abstract}
An inverse scattering problem for the 3D acoustic equation in time domain is
considered. The unknown spatially distributed speed of sound is the subject
of the solution of this problem. A single location of the point source is
used. Using a Carleman Weight Function, a globally strictly convex cost
functional is constructed. A new Carleman estimate is proven. Global
convergence of the gradient projection method is proven. Numerical
experiments are conducted.
\end{abstract}



\section{Introduction}

\label{sec:1}

We develop analytically and test numerically a globally convergent numerical
method for a Coefficient Inverse Problem (CIP) for the 3D acoustic equation
in the time domain. The spatially distributed speed of sound is the subject
of the solution of this CIP. Only a single location of the point source is
used here. For reasons explained below, we call this method
\textquotedblleft convexification". In the publication \cite{Klib97} of the
first author a globally strictly convex cost functional was constructed for
the same CIP for the first time. This paper is a deep revision of \cite%
{Klib97}. We significantly update the theory of \cite{Klib97}. Our goal is
to make the theory of \cite{Klib97} computationally feasible. New results of
this paper are (also, see Remark 4.1 in section 4):

\begin{enumerate}
\item We prove a new Carleman estimate for the Laplace's operator (Theorem
4.1). This one is well suitable for computations. On the other hand, the
Carleman estimate of \cite{Klib97} uses two large parameters and, therefore,
changes too rapidly, which makes it inconvenient for computations.

\item Using the Carleman estimate of Theorem 4.1, we \textquotedblleft
convexify" our CIP, i.e. we construct a new weighted Tikhonov-like
functional, which is strictly convex on a set $B\left( m,R\right) $ in a
certain Hilbert space of an arbitrary \textquotedblleft size" $R>0$, where
the number $m\in \left( 0,R\right) $ is independent on $R$. Since the number 
$R>0$ is an arbitrary one, then this is the \emph{global strict convexity}.
The key element of the above functional is the presence of the Carleman
Weight Function (CWF) in it. Unlike \cite{Klib97}, our functional contains a
regularization parameter $\beta \in \left( 0,1\right) .$ It is the presence
of this parameter which allows us to work without an inconvenient assumption
of \cite{Klib97} (the last paragraph of page 1379 of \cite{Klib97}) which is
about working on a certain compact set.\ Unlike the latter, we consider here
a more convenient case of the bounded set $B\left( m,R\right) $. We call our
approach \textquotedblleft convexification".

\item We prove the existence and uniqueness of the minimizer of that
functional on $B\left( m,R\right) $ and, most importantly, the \emph{global
convergence} of the gradient projection method to the correct solution of
our CIP. In other words, the convergence of the gradient projection method
to the correct solution is guaranteed if it starts from an arbitrary point
of the set $B\left( m,R\right) $ and if the noise in the data tends to zero.

\item This theory is applied then to computational studies of the proposed
numerical method. Such studies are not a part of \cite{Klib97}.
\end{enumerate}

We call a numerical method for a CIP \emph{globally} convergent if a theorem
is proved, which guarantees that this method delivers at least one point in
a sufficiently small neighborhood of the exact solution without any advanced
knowledge of this neighborhood. Thus, since the convergence to the exact
solution of the gradient projection method of the minimization of the above
mentioned weighted Tikhonov-like functional is guaranteed for any starting
point of the set $B\left( m,R\right) $ and since its \textquotedblleft size" 
$R>0$ is an arbitrary number, then the numerical method proposed in this
paper is a globally convergent one.

A conventional numerical methods for a CIP relies on the minimization of a
Tikhonov least squares cost functional, see, e.g. \cite%
{Chav,Gonch,Lakhal,Rizutti}. However, since, as a rule, such a functional is
non convex, then the fundamental and still not addressed problem of such a
method is the presence of multiple local minima and ravines in that
functional. Therefore, convergence of an optimization method for such a
functional can be guaranteed only if its starting point is located in a
sufficiently small neighborhood of the exact solution, i.e. the \emph{local}
convergence. There exists a vast literature about numerical solutions of
inverse scattering problems, which are close to the CIP we consider, see,
e.g. \cite%
{Am1,Am2,Cakoni,Kar,Gonch,It,Ito,Kab1,Kab2,Liu1,Liu2,Lakhal,Rizutti} and
references cited therein. They consider either problems of finding unknown
coefficients or problems of finding shapes of obstacles. In the case of
unknown coefficients, these cited papers consider the case of either
multiple locations of the point source or multiple directions of the
incident plane wave, which is unlike our case of a single point source.

Initially globally strictly convex cost functionals with CWFs in them were
constructed in \cite{Klib95,Klib97,Klib971} for CIPs for hyperbolic and
parabolic PDEs. Numerical studies were not conducted in these works. In the
past few years an interest to this topic was renewed in publications of the
first author with coauthors. New analytical results were combined with
computational ones for some CIPs \cite%
{KlibThanh,KlibKol1,KlibKol2,KlibKol3,KEIT}, including a quite challenging
case of experimental data \cite{KlibKol4}. However, the convexification was
not studied numerically for CIPs in time domain in the case when the initial
condition is vanishing, as we have here. In the case of a non vanishing
initial condition, a different version of the convexification was recently
proposed in \cite{Baud} for the PDE $u_{tt}=\Delta u+a(x)u$ with the unknown
coefficient $a(x)$. The case of a non vanishing initial condition is less
challenging one than our case of the $\delta -$function in the initial
condition. An indication of the latter is that the uniqueness theorem for
the non vanishing case can be proved by the Bukhgeim-Klibanov method, which
was originated in \cite{BukhKlib}.\ On the other hand, it is still unclear
how to prove uniqueness theorem for our case of the $\delta -$function in
the initial condition. Thus, we just assume uniqueness here for
computational purposes. Since this paper is not about the Bukhgeim-Klibanov
method, we cite now, in addition to the above first publication, only some
limited relevant references \cite{BK,Bell,KT,Ksurvey}.

In section 2 we state the CIP we consider. In section 3 we derive a boundary
value problem for a system of coupled elliptic PDEs. In section 4 we derive
the above mentioned weighted globally strictly convex Tikhonov-like
functional with a CWF in it and also formulate our Theorems 4.1-4.6. In
section 5 we prove these theorems. In section 6 we present results of
numerical experiments.

\section{Statement of the Inverse Problem}

\label{sec:2}

Below $\mathbf{x=}\left( x,y,z\right) =\left( x_{1},x_{2},x_{3}\right) \in 
\mathbb{R}^{3}.$ Let $A>0$ be a number. Since the domain of interest $\Omega 
$ is a cube in our computations, then it is convenient to set $\Omega
\subset \mathbb{R}^{3}$ as%
\begin{equation}
\Omega =\left\{ \mathbf{x=}\left( x,y,z\right) :-A/2<x,y<A/2,z\in \left(
0,A\right) \right\} .  \label{2.1}
\end{equation}%
Let the number $a>0.$ We set the single point source we use as $\mathbf{x}%
_{0}=\left( 0,0,-a\right) $. Hence, this source is located below the domain $%
\Omega .$ Let $\Gamma _{0}$ be the upper boundary of $\Omega $ and $\Gamma
_{1}$ be the rest of this boundary, 
\begin{equation}
\Gamma _{0}=\left\{ \mathbf{x=}\left( x,y,z\right) :-A/2<x,y<A/2,z=A\right\}
,\Gamma _{1}=\partial \Omega \diagdown \Gamma _{0}.  \label{2.01}
\end{equation}%
Thus, $\Gamma _{0}$ is the \textquotedblleft transmitted" side of $\Omega .$
Let the function $c\left( \mathbf{x}\right) $ be such that%
\begin{equation}
c\in C^{13}\left( \mathbb{R}^{3}\right) ,  \label{2.2}
\end{equation}%
\begin{equation}
c\left( \mathbf{x}\right) \geq c_{0}=const.>0\text{ in }\overline{\Omega },
\label{2.3}
\end{equation}%
\begin{equation}
c\left( \mathbf{x}\right) =1,\forall \mathbf{x}\in \mathbb{R}^{3}\diagdown
\Omega .  \label{2.4}
\end{equation}

\textbf{Remark 2.1}. \emph{We assume in (\ref{2.1}) the }$C^{13}-$\emph{%
smoothness of the function }$c\left( \mathbf{x}\right) $\emph{\ since the
representation (\ref{2.9}) of the solution of the Cauchy problem (\ref{2.7}%
), (\ref{2.8}) works only under this assumption. Indeed, this smoothness was
carefully calculated in Theorem 4.1 of the book \cite{Rom}.}

The physical meaning of the function $c\left( \mathbf{x}\right) $ is that $1/%
\sqrt{c\left( \mathbf{x}\right) }$ is the speed of sound. Consider the
conformal Riemannian metric generated by the function $c\left( \mathbf{x}%
\right) ,$%
\begin{equation}
d\tau =\sqrt{c\left( \mathbf{x}\right) }\sqrt{\left( dx\right) ^{2}+\left(
dy\right) ^{2}+\left( dz\right) ^{2}}.  \label{2.5}
\end{equation}%
The metric (\ref{2.5}) generates geodesic lines $\Gamma \left( \mathbf{x},%
\mathbf{x}_{0}\right) $, $\mathbf{x}\in \mathbb{R}^{3}$. Let $\tau \left( 
\mathbf{x}\right) $ is the travel time along the geodesic line $\Gamma
\left( \mathbf{x},\mathbf{x}_{0}\right) .$ Then \cite{Rom} the function $%
\tau \left( \mathbf{x}\right) $ is the solution of the eikonal equation%
\begin{equation}
\left\vert \nabla \tau \left( \mathbf{x}\right) \right\vert ^{2}=c\left( 
\mathbf{x}\right) ,  \label{2.6}
\end{equation}%
with the condition $\tau \left( \mathbf{x}\right) =O\left( \left\vert 
\mathbf{x}-\mathbf{x}_{0}\right\vert \right) $ as $\left\vert \mathbf{x}-%
\mathbf{x}_{0}\right\vert \rightarrow 0.$ Furthermore,%
\[
\tau \left( \mathbf{x}\right) =\mathop{\displaystyle \int}\limits_{\Gamma
\left( \mathbf{x},\mathbf{x}_{0}\right) }\sqrt{c\left( \mathbf{y}\right) }%
d\sigma . 
\]%
Everywhere below we rely on the following assumption without further
comments \cite{Rom}:

\textbf{Regularity Assumption}. \emph{For the above specific point source }$%
\mathbf{x}_{0},$ \emph{geodesic lines generated by the function }$c\left( 
\mathbf{x}\right) $\emph{\ are regular. In other words, for any points }$%
\mathbf{x}\in \mathbb{R}^{3}$\emph{\ there exists unique geodesic line }$%
\Gamma \left( \mathbf{x},\mathbf{x}_{0}\right) $\emph{\ connecting points }$%
\mathbf{x}$\emph{\ and }$\mathbf{x}_{0}$\emph{.}

A sufficient condition of the regularity of geodesic lines was derived in 
\cite{Rom2}. This condition is%
\[
\mathop{\displaystyle \sum }\limits_{i,j=1}^{3}\frac{\partial ^{2}\ln
c\left( \mathbf{x}\right) }{\partial x_{i}\partial x_{j}}\xi _{i}\xi
_{j}\geq 0,\text{ }\forall \mathbf{x,\xi }\in \mathbb{R}^{3}. 
\]

As the forward problem, we consider\textbf{\ }the following Cauchy problem
for the acoustic equation \cite{Colton} of the hyperbolic type for the
function $u\left( \mathbf{x},\mathbf{x}_{0},t\right) :$ 
\begin{equation}
c\left( \mathbf{x}\right) u_{tt}=\Delta u,\mathbf{x}\in \mathbb{R}^{3},t>0,
\label{2.7}
\end{equation}%
\begin{equation}
u\left( \mathbf{x},0\right) =0,u_{t}\left( \mathbf{x},0\right) =\delta
\left( \mathbf{x}-\mathbf{x}_{0}\right) .  \label{2.8}
\end{equation}%
It was proven in Theorem 4.1 of \cite{Rom} that, given the above conditions (%
\ref{2.2})-(\ref{2.4}) as well as the Regularity Assumption, the problem (%
\ref{2.7}), (\ref{2.8}) has unique solution $u\left( \mathbf{x},\mathbf{x}%
_{0},t\right) $ which can be represented as 
\begin{equation}
u\left( \mathbf{x},t\right) =A\left( \mathbf{x}\right) \delta \left( t-\tau
\left( \mathbf{x}\right) \right) +H\left( t-\tau \left( \mathbf{x}\right)
\right) \widehat{u}\left( \mathbf{x},t\right) ,  \label{2.9}
\end{equation}%
where $\tau \left( \mathbf{x}\right) ,A\left( \mathbf{x}\right) \in
C^{12}\left( \mathbb{R}^{3}\right) $, the function $\widehat{u}\left( 
\mathbf{x},t\right) \in C^{2}\left( t\geq \tau \left( \mathbf{x}\right)
\right) $ and $A\left( \mathbf{x}\right) >0,\forall \mathbf{x}\in \mathbb{R}%
^{3}.$ In (\ref{2.9}) $H\left( z\right) $ is the Heaviside function,%
\[
H\left( z\right) =\left\{ 
\begin{array}{c}
1,z>0, \\ 
0,z<0.%
\end{array}%
\right. 
\]%
Let the number $T>0$. Denote $S_{T}=\partial \Omega \times \left( 0,T\right) 
$ and $\Gamma _{0,T}=\Gamma _{0}\times \left( 0,T\right) .$ Since we work
with only a single location $\mathbf{x}_{0}$ of the point source, we will
omit below indications of dependencies on $\mathbf{x}_{0}.$

\textbf{Coefficient Inverse Problem (CIP)}. \emph{Let the domain }$\Omega $%
\emph{\ be as in (\ref{2.1}). Suppose that the following two functions are
given:}%
\begin{equation}
u\left( \mathbf{x},t\right) \mid _{\left( \mathbf{x,}t\right) \in
S_{T}}=f_{0}\left( \mathbf{x},t\right) ,\partial _{z}u\left( \mathbf{x}%
,t\right) \mid _{\left( \mathbf{x,}t\right) \in \Gamma _{0,T}}=f_{1}\left( 
\mathbf{x},t\right) ,  \label{2.10}
\end{equation}%
\emph{Determine the function }$c\left( \mathbf{x}\right) $\emph{\ for }$%
\mathbf{x}\in \Omega .$

The knowledge of the normal derivative of the function $u\left( \mathbf{x}%
,t\right) $ at the upper boundary of the cube $\Omega $ can be justified as
follows. Suppose that measurements $\varphi \left( x,y,t\right) $ of the
amplitude $u\left( \mathbf{x},t\right) $ of acoustic waves are conducted on
the surface $\Gamma _{1}$ as well as on the plane $\left\{ z=A\right\} .$
Using (\ref{2.4}), (\ref{2.7}) and (\ref{2.8}), we obtain in the half space $%
\left\{ z>A\right\} $%
\[
u_{tt}=\Delta u,\mathbf{x}\in \left\{ z>A\right\} ,t>0, 
\]%
\[
u\left( \mathbf{x},0\right) =u_{t}\left( \mathbf{x},0\right) =0, 
\]%
\[
u\left( x,y,A\right) =\varphi \left( x,y,t\right) ,\left( x,y\right) \in 
\mathbb{R}^{2},t>0. 
\]%
Solving this initial boundary value problem, we uniquely obtain the function 
$u\left( x,y,z,t\right) $ for $z>A,t>0.$ Next, we obtain $\partial
_{z}u\left( x,y,A,t\right) .$

\section{A System of Coupled Quasilinear Elliptic Equations}

\label{sec:3}

In this section we reduce the CIP (\ref{2.7})-(\ref{2.10}) to the Cauchy
problem for\emph{\ }a system of coupled elliptic PDEs.

\subsection{The function $w\left( \mathbf{x},t\right) $}

\label{sec:3.1}

Integrate equation (\ref{2.7}) twice with respect to $t$ for points $\mathbf{%
x}\in \Omega .$ Hence, we consider the function $p\left( \mathbf{x},t\right)
,$%
\begin{equation}
p\left( \mathbf{x},t\right) =\mathop{\displaystyle \int}\limits_{0}^{t}dy%
\mathop{\displaystyle \int}\limits_{0}^{y}u\left( \mathbf{x},s\right) ds,%
\text{ }\mathbf{x}\in \Omega .  \label{3.1}
\end{equation}%
Hence, $p_{tt}\left( \mathbf{x},t\right) =u\left( \mathbf{x},t\right) $ for $%
\mathbf{x}\in \Omega .$ Since $\mathbf{x}_{0}\notin \overline{\Omega },$
then $\delta \left( \mathbf{x}-\mathbf{x}_{0}\right) =0$ for $\mathbf{x}\in
\Omega .$ Hence, by (\ref{2.8}) 
\begin{equation}
\mathop{\displaystyle \int}\limits_{0}^{t}dy\mathop{\displaystyle \int}%
\limits_{0}^{y}u_{tt}\left( \mathbf{x},s\right) ds=\mathop{\displaystyle
\int}\limits_{0}^{t}\left( u_{t}\left( \mathbf{x},y\right) -u_{t}\left( 
\mathbf{x},0\right) \right) dy=u\left( \mathbf{x},t\right) =p_{tt}\left( 
\mathbf{x},t\right) ,\text{ }\mathbf{x}\in \Omega .  \label{3.2}
\end{equation}%
Next, by (\ref{2.7}) and (\ref{3.1}) 
\begin{equation}
\mathop{\displaystyle \int}\limits_{0}^{t}dy\mathop{\displaystyle \int}%
\limits_{0}^{y}u_{tt}\left( \mathbf{x},s\right) ds=\mathop{\displaystyle
\int}\limits_{0}^{t}dy\mathop{\displaystyle \int}\limits_{0}^{y}\Delta
u\left( \mathbf{x},s\right) ds=\Delta p\left( \mathbf{x},t\right) ,\text{ }%
\mathbf{x}\in \Omega .  \label{3.3}
\end{equation}%
Comparing (\ref{3.2}) and (\ref{3.3}), we obtain 
\begin{equation}
p_{tt}\left( \mathbf{x},t\right) =\Delta p\left( \mathbf{x},t\right) \text{
for }\left( \mathbf{x},t\right) \in \Omega \times \left( 0,T\right) .
\label{3.4}
\end{equation}

Next, by (\ref{2.9}) and (\ref{3.1})%
\begin{equation}
p\left( \mathbf{x},t\right) =A\left( \mathbf{x}\right) \left( t-\tau \left( 
\mathbf{x}\right) \right) H\left( t-\tau \left( \mathbf{x}\right) \right)
+O\left( \left( t-\tau \left( \mathbf{x}\right) \right) ^{2}\right) H\left(
t-\tau ^{0}\left( \mathbf{x}\right) \right) ,  \label{3.5}
\end{equation}%
where%
\begin{equation}
\left\vert O\left( \left( t-\tau \left( \mathbf{x}\right) \right)
^{2}\right) \right\vert \leq B\left( t-\tau \left( \mathbf{x}\right) \right)
^{2}\text{ as }t\rightarrow \tau ^{+}\left( \mathbf{x}\right) ,  \label{3.6}
\end{equation}%
\begin{equation}
\left\vert \partial _{t}O\left( \left( t-\tau \left( \mathbf{x}\right)
\right) ^{2}\right) \right\vert \leq B\left( t-\tau \left( \mathbf{x}\right)
\right) \text{ as }t\rightarrow \tau ^{+}\left( \mathbf{x}\right)
\label{3.7}
\end{equation}%
with a certain constant $B>0$ independent on $\left( \mathbf{x},t\right) \in
\Omega \times \left( 0,T\right) .$

Consider the function $w\left( \mathbf{x},t\right) $ defined as%
\begin{equation}
w\left( \mathbf{x},t\right) =p\left( \mathbf{x},t+\tau \left( \mathbf{x}%
\right) \right) ,\text{ for }\left( \mathbf{x},t\right) \in \Omega \times
\left( 0,T\right) .  \label{3.8}
\end{equation}%
Then it follows from the above that $w\in C^{2}\left( \overline{\Omega }%
\times \left[ 0,T\right] \right) $ and by (\ref{3.5})-(\ref{3.7})%
\begin{equation}
w\left( \mathbf{x},0\right) =0,  \label{3.9}
\end{equation}%
\begin{equation}
w_{t}\left( \mathbf{x},0\right) =A\left( \mathbf{x}\right) >0.  \label{3.10}
\end{equation}%
Substituting (\ref{3.8}) in (\ref{2.7}) and using (\ref{2.6}), we obtain%
\begin{equation}
\Delta w-2\mathop{\displaystyle \sum }\limits_{i=1}^{3}w_{x_{i}t}\tau
_{x_{i}}-w_{t}\Delta \tau =0,\mathbf{x}\in \Omega ,t\in \left(
0,T_{1}\right) ,  \label{3.11}
\end{equation}%
\begin{equation}
T_{1}=T-\max_{\overline{\Omega }}\tau \left( \mathbf{x}\right) .
\label{3.110}
\end{equation}
Denote $\widetilde{f}_{0}\left( \mathbf{x},t\right) =f_{0}\left( \mathbf{x}%
,t+\tau \left( \mathbf{x}\right) \right) $ and $\widetilde{f}_{1}\left( 
\mathbf{x},t\right) =f_{1}\left( \mathbf{x},t+\tau \left( \mathbf{x}\right)
\right) .$ Then by (\ref{2.10})%
\begin{equation}
w\left( \mathbf{x},t\right) \mid _{\left( \mathbf{x,}t\right) \in S_{T_{1}}}=%
\widetilde{f}_{0}\left( \mathbf{x},t\right) ,\text{ }\partial _{z}w\left( 
\mathbf{x},t\right) \mid _{\left( \mathbf{x,}t\right) \in \Gamma _{0,T_{1}}}=%
\widetilde{f}_{1}\left( \mathbf{x},t\right) .  \label{3.12}
\end{equation}

Thus, our goal below is to construct a numerical method, which would
approximately find the functions $w\left( \mathbf{x},t\right) ,\tau \left( 
\mathbf{x}\right) $ for $\mathbf{x}\in \Omega ,t\in \left( 0,T_{1}\right) $
from conditions (\ref{3.9})-(\ref{3.12}). Suppose that these two functions
are approximated. Then the corresponding approximation for the target
coefficient $c\left( \mathbf{x}\right) $ can be easily found via the
backwards calculation,

\begin{equation}
c\left( \mathbf{x}\right) =\left\vert \nabla \tau \left( \mathbf{x}\right)
\right\vert ^{2}.  \label{3.120}
\end{equation}%
%
%
%

\subsection{The system of coupled quasilinear elliptic PDEs}

\label{sec:3.2}

\textbf{Lemma 3.1}. \emph{Consider the set of functions }%
\begin{equation}
\left\{ t,t^{2},...,t^{n},...\right\} =\left\{ t^{n}\right\} _{n=1}^{\infty
}.  \label{3.13}
\end{equation}%
\emph{\ Then this set is complete in }$L_{2}\left( 0,T_{1}\right) .$

\textbf{Proof}. Let a function $f\left( t\right) \in L_{2}\left(
0,T_{1}\right) $ be such that%
\[
\mathop{\displaystyle \int}\limits_{0}^{T_{1}}f\left( t\right)
t^{n}dt=0,n=1,2,... 
\]%
Consider the function $\widetilde{f}\left( t\right) =f\left( t\right) t.$
Then 
\begin{equation}
\mathop{\displaystyle \int}\limits_{0}^{T_{1}}\widetilde{f}\left( t\right)
t^{m}dt=0,m=0,1,2,...  \label{3.14}
\end{equation}%
It is well known that (\ref{3.14}) implies that $\widetilde{f}\left(
t\right) \equiv 0.$ $\square $

Orthonormalize the set (\ref{3.13}) using the Gram-Schmidt
orthonormalization procedure. Then Lemma 3.1 implies that we obtain a basis $%
\left\{ P_{n}\left( t\right) \right\} _{n=1}^{\infty }$ of orthonormal
polynomials in $L_{2}\left( 0,T_{1}\right) $ such that 
\begin{equation}
P_{n}\left( 0\right) =0,\forall n=1,2,...  \label{3.15}
\end{equation}%
By (\ref{3.15}) this is not a set of standard orthonormal polynomials.

Let the integer $N>1.$ Approximate the function $w\left( \mathbf{x},t\right) 
$ as%
\begin{equation}
w\left( \mathbf{x},t\right) =\mathop{\displaystyle \sum }%
\limits_{n=1}^{N}w_{n}\left( \mathbf{x}\right) P_{n}\left( t\right) .
\label{3.16}
\end{equation}%
Here and below we use \textquotedblleft $=$" instead of \textquotedblleft $%
\approx "$ for convenience. Substitute (\ref{3.16}) in the left hand side of
(\ref{3.11}) and assume that the resulting left hand side equals zero. We
obtain for $\mathbf{x}\in \Omega $ 
\begin{equation}
\mathop{\displaystyle \sum }\limits_{n=1}^{N}\Delta w_{n}\left( \mathbf{x}%
\right) P_{n}\left( t\right) -2\mathop{\displaystyle \sum }%
\limits_{i=1}^{3}\tau _{x_{i}}\left( \mathbf{x}\right) \mathop{\displaystyle
\sum }\limits_{n=1}^{N}P_{n}^{\prime }\left( t\right) \partial
_{x_{i}}w_{n}\left( \mathbf{x}\right) -\Delta \tau \left( \mathbf{x}\right) %
\mathop{\displaystyle \sum }\limits_{n=1}^{N}P_{n}^{\prime }\left( t\right)
w_{n}\left( \mathbf{x}\right) =0.  \label{3.17}
\end{equation}%
By (\ref{3.10}) and (\ref{3.16}) we can assume that 
\begin{equation}
\mathop{\displaystyle \sum }\limits_{n=1}^{N}P_{n}^{\prime }\left( 0\right)
w_{n}\left( \mathbf{x}\right) =A\left( \mathbf{x}\right) >0,\text{ }\forall 
\mathbf{x}\in \overline{\Omega }.  \label{3.18}
\end{equation}%
Set in (\ref{2.19}) $t=0.$ Hence, we obtain the first elliptic equation,%
\begin{equation}
\Delta \tau \left( \mathbf{x}\right) +2\left[ \mathop{\displaystyle \sum }%
\limits_{i=1}^{3}\tau _{x_{i}}\mathop{\displaystyle \sum }%
\limits_{n=1}^{N}P_{n}^{\prime }\left( 0\right) \partial _{x_{i}}w_{n}\left( 
\mathbf{x}\right) \right] \left[ \mathop{\displaystyle \sum }%
\limits_{n=1}^{N}P_{n}^{\prime }\left( 0\right) w_{n}\left( \mathbf{x}%
\right) \right] ^{-1}=0,\text{ }\mathbf{x}\in \Omega .  \label{3.180}
\end{equation}%
We rewrite this equation as%
\begin{equation}
\Delta \tau =F_{1}\left( \nabla \tau ,\nabla \widetilde{W},\widetilde{W}%
\right) ,\mathbf{x}\in \Omega ,  \label{3.19}
\end{equation}%
where $\widetilde{W}\left( x\right) =\left( w_{1}\left( \mathbf{x}\right)
,...,w_{N}\left( \mathbf{x}\right) \right) ^{T}.$ Next, for $n=1,...,N$
multiply both sides of (\ref{3.17}) by $P_{n}\left( t\right) $ and integrate
with respect to $t\in \left( 0,T_{1}\right) .$ Replace in the resulting
equation $\Delta \tau $ with the right hand side of (\ref{3.19}). We obtain 
\begin{equation}
\Delta \widetilde{W}=F_{2}\left( \nabla \tau ,\nabla \widetilde{W},%
\widetilde{W}\right) ,\mathbf{x}\in \Omega .  \label{3.20}
\end{equation}

Consider the $\left( N+1\right) -$dimensional vector function 
\begin{equation}
W\left( \mathbf{x}\right) =\left( \tau \left( \mathbf{x}\right) ,\widetilde{W%
}\left( \mathbf{x}\right) \right) ^{T}.  \label{3.200}
\end{equation}
Thus, (\ref{3.12}), (\ref{3.19}) and (\ref{3.20}) lead to the following
Cauchy problem for a system of coupled quasilinear elliptic equations%
\begin{equation}
\Delta W+F\left( \nabla W,W\right) =0,\mathbf{x}\in \Omega ,  \label{3.22}
\end{equation}%
\begin{equation}
W\mid _{\partial \Omega }=q^{0}\left( \mathbf{x}\right) ,\partial _{z}W\mid
_{\Gamma _{0}}=q^{1}\left( \mathbf{x}\right) ,  \label{3.23}
\end{equation}%
\begin{equation}
q^{0}\left( \mathbf{x}\right) =\left( \tau \left( \mathbf{x}\right)
,q_{1}^{0}\left( \mathbf{x}\right) ,...,q_{N}^{0}\left( \mathbf{x}\right)
\right) ^{T},\text{ }q_{n}^{0}\left( \mathbf{x}\right) =\mathop{%
\displaystyle \int}\limits_{0}^{T_{1}}\widetilde{f}_{0}\left( \mathbf{x}%
,t\right) P_{n}\left( t\right) dt,\text{ }\mathbf{x}\in \partial \Omega ,
\label{3.24}
\end{equation}%
\begin{equation}
q^{1}\left( \mathbf{x}\right) =\left( \partial _{z}\tau \left( \mathbf{x}%
\right) ,q_{1}^{1}\left( \mathbf{x}\right) ,...,q_{N}^{1}\left( \mathbf{x}%
\right) \right) ^{T},q_{n}^{1}\left( \mathbf{x}\right) =\mathop{%
\displaystyle \int}\limits_{0}^{T_{1}}\widetilde{f}_{1}\left( \mathbf{x}%
,t\right) P_{n}\left( t\right) dt,\text{ }\mathbf{x}\in \Gamma _{0}.
\label{3.25}
\end{equation}

In (\ref{3.24}) and (\ref{3.25}) $n=1,...,N.$ In (\ref{3.22}) the $\left(
N+1\right) -$dimensional vector function $F\in C^{1}\left( \mathbb{R}%
^{3N+5}\right) $. Thus, we have obtained the system (\ref{3.22}) of coupled
quasilinear elliptic PDEs with the Cauchy data (\ref{3.23})-(\ref{3.25}).
Unknowns in this problem are the function $\tau \left( \mathbf{x}\right) $
and Fourier coefficients $w_{n}\left( \mathbf{x}\right) $ of the function $%
w\left( \mathbf{x},t\right) $ in (\ref{3.16}). Therefore, we solve below the
problem (\ref{3.22})-(\ref{3.25}) of finding the $\left( N+1\right) -$%
dimensional vector function $W\in C^{2}\left( \overline{\Omega }\right) .$
In fact, however, we find below $W\in H^{3}\left( \Omega \right) .$

\textbf{Remarks 3.1}:

\begin{enumerate}
\item \emph{The number }$N$\emph{\ in (\ref{3.16}) should be chosen
computationally, see section 6. Since (\ref{3.16}) is an approximation of
the function }$w\left( \mathbf{x},t\right) ,$\emph{\ then (\ref{3.17})-(\ref%
{3.23}) are also understood in the approximate sense. Thus, (\ref{3.16})-(%
\ref{3.23}) form our \textbf{approximate mathematical model}. Since the
original CIP is a very challenging one, it is hard to anticipate that it
would be effectively solved numerically without such approximations. Since
we develop a numerical method here, then it is fine to introduce an
approximate mathematical model.}

\item \emph{We cannot prove convergence of our resulting solutions to the
correct one as }$N\rightarrow \infty .$\emph{\ The underlying reason of this
is the ill-posed nature of our CIP combined with the nonlinearity. On the
other hand, truncated Fourier series are considered quite often in the field
of Inverse and Ill-Posed Problems, whereas convergencies of resulting
solutions at }$N\rightarrow \infty $\emph{\ are proven only very rarely.
Nevertheless, corresponding approximate mathematical models usually provide
quite good numerical results: we refer, e.g. to \cite%
{Kab1,Kab2,KlibKol3,KEIT} for corresponding publications. }
\end{enumerate}

\section{Globally Strictly Convex Tikhonov-like Functional}

\label{sec:4}

\subsection{The functional}

\label{sec:4.1}

All Banach spaces considered below are spaces of real valued functions. If
we say below that a vector function belongs to a certain Banach space, then
this means that all its components belong to this space, and the norm of
this function in that space is defined as the square root of the sum of
squares of norms of its components.

To arrange a certain projection operator for the gradient projection method
below, the best way is to have zero Cauchy data.\ Hence, we assume that
there exists an $\left( N+1\right) -$dimensional vector function $G=\left(
g_{0}\left( \mathbf{x}\right) ,...,g_{N}\left( \mathbf{x}\right) \right)
^{T}\in H^{3}\left( \Omega \right) $ satisfying boundary conditions (\ref%
{3.23}), i.e. such that 
\begin{equation}
G\mid _{\partial \Omega }=q^{0}\left( \mathbf{x}\right) ,\partial _{z}G\mid
_{\Gamma _{0}}=q^{1}\left( \mathbf{x}\right) .  \label{4.1}
\end{equation}%
Let 
\begin{equation}
W-G=Q\in H^{3}\left( \Omega \right) ,Q\left( \mathbf{x}\right) =\left(
q_{0}\left( \mathbf{x}\right) ,...,q_{N}\left( \mathbf{x}\right) \right)
^{T}.  \label{4.100}
\end{equation}
Then (\ref{3.22}), (\ref{3.23}) and (\ref{4.1}) imply that 
\begin{equation}
\Delta Q+\Delta G+F\left( Q+G,\nabla \left( Q+G\right) \right) =0,\mathbf{x}%
\in \Omega ,  \label{4.2}
\end{equation}%
\begin{equation}
Q\mid _{\partial \Omega }=0,\partial _{z}Q\mid _{\Gamma _{0}}=0.  \label{4.3}
\end{equation}

Let $H_{0}^{3}\left( \Omega \right) $ be the subspace of the space $%
H^{3}\left( \Omega \right) $ defined as%
\[
H_{0}^{3}\left( \Omega \right) =\left\{ v\in H^{3}\left( \Omega \right)
:v\mid _{\partial \Omega }=0,\partial _{z}v\mid _{\Gamma _{0}}=0\right\} . 
\]%
Choose an arbitrary number $R>0$ and also choose another number $m\in \left(
0,R\right) ,$ which is independent on $R$. Consider the set $B\left(
m,R\right) $ of $\left( N+1\right) -$dimensional vector functions $Z\left( 
\mathbf{x}\right) =\left( z_{0}\left( \mathbf{x}\right) ,...,z_{N}\left( 
\mathbf{x}\right) \right) ^{T}$ such that%
\begin{equation}
B\left( m,R\right) =\left\{ 
\begin{array}{c}
Z\in H_{0}^{3}\left( \Omega \right) ,\text{ }\left\Vert Z\right\Vert
_{H^{3}\left( \Omega \right) }<R, \\ 
\mathop{\displaystyle \sum }\limits_{n=1}^{N}P_{n}^{\prime }\left( 0\right)
\left( z_{n}\left( \mathbf{x}\right) +g_{n}\left( \mathbf{x}\right) \right)
>m,\forall \mathbf{x}\in \overline{\Omega }.%
\end{array}%
\right.  \label{4.4}
\end{equation}

The second condition (\ref{4.3}) is generated by (\ref{3.18}). By embedding
theorem \ $H^{3}\left( \Omega \right) \subset C^{1}\left( \overline{\Omega }%
\right) .$ This implies that $\overline{B\left( m,R\right) }\subset
C^{1}\left( \overline{\Omega }\right) $ and also that there exist numbers $%
D_{1}\left( R\right) >0$ and $D_{2}\left( G\right) >0$ depending only on
listed parameters such that 
\begin{equation}
\left\Vert Z\right\Vert _{C^{1}\left( \overline{\Omega }\right) }\leq
D_{1}\left( R\right) ,\text{ }\forall Z\in \overline{B\left( m,R\right) },
\label{4.40}
\end{equation}%
\begin{equation}
\left\Vert G\right\Vert _{C^{1}\left( \overline{\Omega }\right) }\leq
D_{2}\left( G\right) .  \label{4.400}
\end{equation}%
Temporary replace the vector functions $Q\left( \mathbf{x}\right) =$ $\left(
q_{0}\left( \mathbf{x}\right) ,...,q_{N}\left( \mathbf{x}\right) \right)
^{T} $ and $\nabla Q\left( \mathbf{x}\right) =\left( \nabla q_{0}\left( 
\mathbf{x}\right) ,...,\nabla q_{N}\left( \mathbf{x}\right) \right) ^{T}$
with the vector of real variables $\left( y_{0},y_{1},...,y_{4N+3}\right)
^{T}=y\in \mathbb{R}^{4N+4}.$ Consider the set $Y\subset \mathbb{R}^{4N+4}$, 
\[
Y=\left\{ y\in \mathbb{R}^{4N+4}:\mathop{\displaystyle \sum }%
\limits_{n=1}^{N}P_{n}^{\prime }\left( 0\right) \left( y_{n}+g_{n-1}\left( 
\mathbf{x}\right) \right) >m,\forall \mathbf{x}\in \overline{\Omega }%
\right\} . 
\]%
Obviously $Y$ is an open set in $\mathbb{R}^{4N+4}$. Denote $p_{1}=\left(
y_{0},...,y_{N}\right) ,p_{2}=\left(
y_{0,1},y_{0,2},y_{0,3},y_{1,1},...,y_{N,3}\right) .$ Then $y=\left(
p_{1},p_{2}\right) ^{T}\in \mathbb{R}^{4N+4}.$ It follows from (\ref{3.180}%
)-(\ref{3.22}) that, as the function of $y,$ 
\begin{equation}
F\left( p_{1}+G\left( \mathbf{x}\right) ,p_{2}+\nabla G\left( \mathbf{x}%
\right) \right) \in C^{2}\left( \overline{Y}\right) ,\forall \mathbf{x}\in 
\overline{\Omega }.  \label{4.42}
\end{equation}

\textbf{Lemma 4.1}. \emph{The set }$B\left( m,R\right) $\emph{\ is convex.}

\textbf{Proof}. Let the number $\alpha \in \left( 0,1\right) $ and vector
functions $Z,Y\in B\left( m,R\right) .$ Consider the function $\alpha
Z+\left( 1-\alpha \right) Y.$ Then 
\[
\left\Vert \alpha Z+\left( 1-\alpha \right) Y\right\Vert _{H^{3}\left(
\Omega \right) }\leq \alpha \left\Vert Z\right\Vert _{H^{3}\left( \Omega
\right) }+\left( 1-\alpha \right) \left\Vert Y\right\Vert _{H^{3}\left(
\Omega \right) }<\alpha R+\left( 1-\alpha \right) R=R. 
\]%
Next, let $Z\left( \mathbf{x}\right) =\left( \tau _{1}\left( \mathbf{x}%
\right) ,z_{1}\left( \mathbf{x}\right) ,...,z_{N}\left( \mathbf{x}\right)
\right) ^{T},Y\left( \mathbf{x}\right) =\left( \tau _{2}\left( \mathbf{x}%
\right) ,y_{1}\left( \mathbf{x}\right) ,...,y_{N}\left( \mathbf{x}\right)
\right) ^{T}.$ Then 
\[
\alpha \mathop{\displaystyle \sum }\limits_{n=1}^{N}P_{n}^{\prime }\left(
0\right) z_{n}\left( \mathbf{x}\right) +\left( 1-\alpha \right) %
\mathop{\displaystyle \sum }\limits_{n=1}^{N}P_{n}^{\prime }\left( 0\right)
y_{n}\left( \mathbf{x}\right) >\alpha m+\left( 1-\alpha \right) m=m.\text{ \ 
}\square 
\]

Our weighted Tikhonov-like cost functional is%
\begin{equation}
J_{\lambda ,\beta }\left( Q+G\right) =e^{-2\lambda b^{2}}\mathop{%
\displaystyle \int}\limits_{\Omega }\left( \Delta Q+\Delta G+F\left( \nabla
\left( Q+G\right) ,Q+G\right) \right) ^{2}e^{2\lambda \left( z+b\right)
^{2}}d\mathbf{x}  \label{4.5}
\end{equation}%
\[
+\beta \left\Vert Q+G\right\Vert _{H^{3}\left( \Omega \right) }^{2}. 
\]%
In (\ref{4.5}) the numbers $\lambda \geq 1,b>0,\beta \in \left( 0,1\right) .$
Here, $\lambda $ is the parameter of our CWF $e^{2\lambda \left( z+b\right)
^{2}}$ and $\beta $ is the regularization parameter. The multiplier $%
e^{-2\lambda b^{2}}$ is introduced to balance two terms in the right hand
side of (\ref{4.5}). Indeed, by (\ref{2.1}) 
\begin{equation}
\min_{\mathbf{x}\in \overline{\Omega }}\left( e^{-2\lambda b^{2}}e^{2\lambda
\left( z+b\right) ^{2}}\right) =1.  \label{4.50}
\end{equation}

\textbf{Minimization Problem. }\emph{Minimize the functional }$J_{\lambda
,\beta }\left( Q\right) $\emph{\ in (\ref{4.5}) on the set }$B\left(
m,R\right) $\emph{\ defined in (\ref{4.4}).}

\subsection{Theorems}

\label{sec:4.2}

\textbf{Theorem 4.1}. \emph{There exists a sufficiently large number }$%
\lambda _{0}=\lambda _{0}\left( \Omega ,b\right) \geq 1$\emph{\ and a
constant }$C_{1}=C_{1}\left( \Omega ,b\right) >0,$\emph{\ both depending
only on }$\Omega $\emph{\ and }$b,$\emph{\ such that for all }$\lambda \geq
\lambda _{0}$\emph{\ and for all functions }$u\in H^{2}\left( \Omega \right) 
$\emph{\ such that }$u\mid _{\partial \Omega }=u_{z}\mid _{\Gamma _{0}}=0$%
\emph{\ the following Carleman estimate holds}%
\begin{equation}
\mathop{\displaystyle \int}\limits_{\Omega }\left( \Delta u\right)
^{2}e^{2\lambda \left( z+b\right) ^{2}}d\mathbf{x}\geq \frac{C_{1}}{\lambda }%
\mathop{\displaystyle \sum }\limits_{i,j=1}^{3}\mathop{\displaystyle \int}%
\limits_{\Omega }u_{x_{i}x_{j}}^{2}e^{2\lambda \left( z+b\right) ^{2}}d%
\mathbf{x}  \label{4.6}
\end{equation}%
\[
+C_{1}\lambda \mathop{\displaystyle \int}\limits_{\Omega }\left( \left(
\nabla u\right) ^{2}+\lambda ^{2}u^{2}\right) e^{2\lambda \left( z+b\right)
^{2}}d\mathbf{x.} 
\]

Below $C_{2}=C_{2}\left( F,\left\Vert G\right\Vert _{H^{3}\left( \Omega
\right) },m,R,\Omega ,b\right) >0$ denotes different constants depending
only on listed parameters.

\textbf{Theorem 4.2 }(global strict convexity). \emph{For all }$Q\in B\left(
m,2R\right) $\emph{, }$\lambda ,\beta >0$\emph{\ there exists Frech\'{e}t
derivative }$J_{\lambda ,\beta }^{\prime }\left( Q+G\right) \in
H_{0}^{3}\left( \Omega \right) .$\emph{\ Let }$\lambda _{0}$\emph{\ be the
number of Theorem 4.1. There exists a number }$\lambda _{1}=\lambda
_{1}\left( F,\left\Vert G\right\Vert _{H^{3}\left( \Omega \right)
},m,R,\Omega ,b\right) \geq \lambda _{0}$\emph{\ depending only on listed
parameters such that for any }$\lambda \geq \lambda _{1}$\emph{\ and any }$%
\beta >0$\emph{\ the functional }$J_{\lambda ,\beta }\left( Q\right) $\emph{%
\ is strictly convex on }$B\left( m,R\right) ,$\emph{\ i.e. the following
estimate holds for all }$Q_{1},Q_{2}\in B\left( m,R\right) $%
\[
J_{\lambda ,\beta }\left( Q_{2}+G\right) -J_{\lambda ,\beta }\left(
Q_{1}+G\right) -J_{\lambda ,\beta }^{\prime }\left( Q_{1}+G\right) \left(
Q_{2}-Q_{1}\right) 
\]%
\begin{equation}
\geq \frac{C_{2}}{\lambda }\mathop{\displaystyle \sum }\limits_{i,j=1}^{3}%
\left\Vert \left( Q_{2}-Q_{1}\right) _{x_{i}x_{j}}\right\Vert _{L_{2}\left(
\Omega \right) }^{2}+C_{2}\lambda \left\Vert Q_{2}-Q_{1}\right\Vert
_{H^{1}\left( \Omega \right) }^{2}+\beta \left\Vert Q_{2}-Q_{1}\right\Vert
_{H^{3}\left( \Omega \right) }^{2}.  \label{4.7}
\end{equation}

\textbf{Theorem 4.3}. \emph{The} \emph{Frech\'{e}t derivative }$J_{\lambda
,\beta }^{\prime }\left( Q+G\right) \in H_{0}^{3}\left( \Omega \right) $ 
\emph{of the functional }$J_{\lambda ,\beta }\left( Q\right) $\emph{\
satisfies the Lipschitz continuity condition in }$B\left( m,2R\right) $\emph{%
\ for all }$\lambda ,\beta >0.$\emph{\ In other words, there exists a number 
}$L=L\left( F,\left\Vert G\right\Vert _{H^{3}\left( \Omega \right)
},m,R,\Omega ,b,\lambda ,\beta \right) $\emph{\ depending only on listed
parameters such that} 
\[
\left\Vert J_{\lambda ,\beta }^{\prime }\left( Q_{2}+G\right) -J_{\lambda
,\beta }^{\prime }\left( Q_{1}+G\right) \right\Vert _{H^{3}\left( \Omega
\right) }\leq L\left\Vert Q_{2}-Q_{1}\right\Vert _{H^{3}\left( \Omega
\right) },\text{ }\forall Q_{1},Q_{2}\in B\left( m,2R\right) . 
\]

\textbf{Theorem 4.4}. \emph{For each pair }$\lambda \geq $\emph{\ }$\lambda
_{1},\beta >0$\emph{\ there exists unique minimizer }$Q_{\min ,\lambda
,\beta }\in \overline{B\left( m,R\right) }$\emph{\ of the functional }$%
J_{\lambda ,\beta }\left( Q\right) $\emph{\ on the set }$\overline{B\left(
m,R\right) }.$\emph{\ Furthermore,} 
\begin{equation}
J_{\lambda ,\beta }^{\prime }\left( Q_{\min ,\lambda ,\beta }+G\right)
\left( Q_{\min ,\lambda ,\beta }-p\right) \leq 0,\text{ }\forall p\in
H_{0}^{3}\left( \Omega \right) .  \label{4.8}
\end{equation}

We now arrange the gradient projection method of the minimization of the
functional $J_{\lambda ,\beta }\left( Q+G\right) $ on the set $\overline{%
B\left( m,R\right) }.$ Let the number $\gamma \in \left( 0,1\right) .$ Let $%
P_{\overline{B}}:H_{0}^{3}\left( \Omega \right) \rightarrow \overline{%
B\left( m,R\right) }$ \ be the projection operator of the space $H_{0}^{3}$
on the set $\overline{B\left( m,R\right) }.$\ Let $Q_{0}\in B\left(
m,R\right) $ be an arbitrary point of this set. The gradient projection
method amounts to the following sequence:%
\begin{equation}
Q_{n}=P_{\overline{B}}\left( Q_{n-1}-\gamma J_{\lambda ,\beta }^{\prime
}\left( Q_{n-1}+G\right) \right) ,n=1,2,...  \label{4.9}
\end{equation}

\textbf{Theorem 4.5}. \emph{Let }$\lambda _{1}$\emph{\ and }$\beta $\emph{\
be parameters of Theorem 4.2. Choose a number }$\lambda \geq \lambda _{1}.$%
\emph{\ Let }$Q_{\min ,\lambda ,\beta }\in \overline{B\left( m,R\right) }$%
\emph{\ be the unique minimizer of the functional }$J_{\lambda ,\beta
}\left( Q\right) $\emph{\ on the set }$\overline{B\left( m,R\right) }$\emph{%
\ (Theorem 4.3). Then there exists a sufficiently small number }$\gamma
_{0}=\gamma _{0}\left( F,\left\Vert G\right\Vert _{H^{3}\left( \Omega
\right) },m,R,\Omega ,b,\beta \right) \in \left( 0,1\right) $\emph{\
depending only on listed parameters such that the sequence (\ref{4.9})
converges to }$Q_{\min ,\lambda ,\beta }$\emph{\ in the space }$H^{3}\left(
\Omega \right) .$\emph{\ More precisely, there exists a number }$\theta
=\theta \left( F,\left\Vert G\right\Vert _{H^{3}\left( \Omega \right)
},m,R,\Omega ,b,\beta \right) \in \left( 0,1\right) $\emph{\ such that the
following estimate holds}%
\begin{equation}
\left\Vert Q_{n}-Q_{\min ,\lambda ,\beta }\right\Vert _{H^{3}\left( \Omega
\right) }\leq \theta ^{n}\left\Vert Q_{\min ,\lambda ,\beta
}-Q_{0}\right\Vert _{H^{3}\left( \Omega \right) }.  \label{4.90}
\end{equation}

Following the Tikhonov concept for ill-posed problems \cite{BK,T}, we now
assume the existence of the exact solution $Q^{\ast }\left( \mathbf{x}%
\right) =\left( q_{0}^{\ast }\left( \mathbf{x}\right) ,...,q_{N}^{\ast
}\left( \mathbf{x}\right) \right) ^{T}\in B\left( m,R\right) $ of the
problem (\ref{4.2}), (\ref{4.3}) with the noiseless data $G^{\ast }\left( 
\mathbf{x}\right) =\left( g_{0}^{\ast }\left( \mathbf{x}\right)
,...,g_{N}^{\ast }\left( \mathbf{x}\right) \right) ^{T}\in H^{3}\left(
\Omega \right) .$ In particular, this means that 
\[
\mathop{\displaystyle \sum }\limits_{n=1}^{N}P_{n}^{\prime }\left( 0\right)
\left( q_{n}^{\ast }\left( \mathbf{x}\right) +g_{n}^{\ast }\left( \mathbf{x}%
\right) \right) >m,\forall \mathbf{x}\in \overline{\Omega }. 
\]%
Also, let the number $\delta \in \left( 0,1\right) $ be the level of the
error in the data $G,$ i.e.%
\begin{equation}
\left\Vert G-G^{\ast }\right\Vert _{H^{3}\left( \Omega \right) }<\delta .
\label{4.10}
\end{equation}%
Since $\delta \in \left( 0,1\right) ,$ then (\ref{4.10}) implies that we can
regard in Theorem 4.6 that constants $\lambda _{1},C_{2},\gamma _{0},\theta $
introduced above depend on $\left\Vert G^{\ast }\right\Vert _{H^{3}\left(
\Omega \right) }$ rather than on $\left\Vert G\right\Vert _{H^{3}\left(
\Omega \right) }.$ We are doing so both in the formulation and in the proof
of Theorem 4.6.

\textbf{Theorem 4.6 }(error estimates and convergence).\textbf{\ }\emph{Let }%
$\lambda _{1}=\lambda _{1}\left( F,\left\Vert G\right\Vert _{H^{3}\left(
\Omega \right) },m,R,\Omega ,b\right) $\emph{\ be the number of Theorem 4.2.
Define the number }$\eta $\emph{\ as }$\eta =\left[ 4\left( A+b\right) ^{2}%
\right] ^{-1}.$\emph{\ Choose a sufficiently small number }$\delta _{0}\in
\left( 0,1\right) $\emph{\ such that }$\ln \delta _{0}^{-\eta }\geq \lambda
_{1}.$\emph{\ Let in (\ref{4.10}) }$\delta \in \left( 0,\delta _{0}\right) $%
\emph{. Choose }$\lambda =\lambda \left( \delta \right) =\ln \delta ^{-\eta
}>\lambda _{1}$ \emph{implying that} $\exp \left[ 2\lambda \left( \delta
\right) \left( A+b\right) ^{2}\right] =1/\sqrt{\delta }.$ \emph{Let }$%
Q_{\min ,\lambda ,\beta }\in $\emph{\ }$\overline{B\left( m,R\right) }$\emph{%
\ be the unique minimizer of the functional }$J_{\lambda ,\beta }\left(
Q\right) $\emph{\ on the set }$\overline{B\left( m,R\right) }$\emph{, the
existence of which is guaranteed by Theorem 4.4. Let }$\left\{ Q_{n}\right\}
_{n=0}^{\infty }$\emph{\ }$\subset \overline{B\left( m,R\right) }$\emph{\ be
the sequence of the gradient projection method (\ref{4.9}) with an arbitrary
starting point }$Q_{0}\in \overline{B\left( m,R\right) }.$\emph{\ Then the
following estimates hold for all }$\beta \in \left( 0,1\right) $%
\begin{equation}
\left\Vert Q^{\ast }-Q_{\min ,\lambda ,\beta }\right\Vert _{H^{1}\left(
\Omega \right) }\leq C_{2}\left( \delta ^{\eta /2+1/4}+\sqrt{\beta }\delta
^{\eta /2}\right) ,  \label{1}
\end{equation}%
\begin{equation}
\left\Vert Q^{\ast }-Q_{\min ,\lambda ,\beta }\right\Vert _{H^{2}\left(
\Omega \right) }\leq C_{2}\left( \delta ^{1/4}+\sqrt{\beta }\right) \sqrt{%
\ln \delta ^{-\eta }},  \label{2}
\end{equation}%
\begin{equation}
\left\Vert Q^{\ast }-Q_{n}\right\Vert _{H^{1}\left( \Omega \right) }\leq
C_{2}\left( \delta ^{\eta /2+1/4}+\sqrt{\beta }\delta ^{\eta /2}\right)
+\theta ^{n}\left\Vert Q_{\min ,\lambda ,\beta }-Q_{0}\right\Vert
_{H^{3}\left( \Omega \right) },  \label{3}
\end{equation}%
\begin{equation}
\left\Vert Q^{\ast }-Q_{n}\right\Vert _{H^{2}\left( \Omega \right) }\leq
C_{2}\left( \delta ^{1/4}+\sqrt{\beta }\right) \sqrt{\ln \delta ^{-\eta }}%
+\theta ^{n}\left\Vert Q_{\min ,\lambda ,\beta }-Q_{0}\right\Vert
_{H^{3}\left( \Omega \right) },  \label{4}
\end{equation}%
\begin{equation}
\left\Vert c^{\ast }-c_{n}\right\Vert _{L_{2}\left( \Omega \right) }\leq
C_{2}\left( \delta ^{1/4}+\sqrt{\beta }\right) \sqrt{\ln \delta ^{-\eta }}%
+\theta ^{n}\left\Vert Q_{\min ,\lambda ,\beta }-Q_{0}\right\Vert
_{H^{3}\left( \Omega \right) },  \label{5}
\end{equation}%
\emph{In particular, if the regularization parameter }$\beta =\sqrt{\delta }$%
\emph{, as required by the regularization theory \cite{T}, then estimates (%
\ref{1})-(\ref{4}) become} 
\[
\left\Vert Q^{\ast }-Q_{\min ,\lambda ,\beta }\right\Vert _{H^{1}\left(
\Omega \right) }\leq C_{2}\delta ^{\eta /2+1/4} 
\]%
\[
\left\Vert Q^{\ast }-Q_{\min ,\lambda ,\beta }\right\Vert _{H^{2}\left(
\Omega \right) }\leq C_{2}\delta ^{1/4}\sqrt{\ln \delta ^{-\eta }}, 
\]%
\[
\left\Vert Q^{\ast }-Q_{n}\right\Vert _{H^{1}\left( \Omega \right) }\leq
C_{2}\delta ^{\eta /2+1/4}+\theta ^{n}\left\Vert Q_{\min ,\lambda ,\sqrt{%
\delta }}-Q_{0}\right\Vert _{H^{3}\left( \Omega \right) }, 
\]%
\[
\left\Vert Q^{\ast }-Q_{n}\right\Vert _{H^{2}\left( \Omega \right) }\leq
C_{2}\delta ^{1/4}\sqrt{\ln \delta ^{-\eta }}+\theta ^{n}\left\Vert Q_{\min
,\lambda ,\sqrt{\delta }}-Q_{0}\right\Vert _{H^{3}\left( \Omega \right) }, 
\]%
\[
\left\Vert c^{\ast }-c_{n}\right\Vert _{L_{2}\left( \Omega \right) }\leq
C_{2}\delta ^{1/4}\sqrt{\ln \delta ^{-\eta }}+\theta ^{n}\left\Vert Q_{\min
,\lambda ,\sqrt{\delta }}-Q_{0}\right\Vert _{H^{3}\left( \Omega \right) }. 
\]%
\emph{Here} $c_{n}\left( \mathbf{x}\right) $ \emph{is defined via (\ref%
{4.100}) with }$Q=Q_{n}$\emph{, next (\ref{3.200}), and next (\ref{3.120}).
Further,} $c^{\ast }\left( \mathbf{x}\right) $ \emph{is defined as the exact
target coefficient, which corresponds to the noiseless data }$G^{\ast }$%
\emph{\ with }$Q^{\ast }$\emph{\ and }$W^{\ast }=Q^{\ast }+G^{\ast }$\emph{\
in (\ref{4.100}), and next similarly as for} $c_{n}\left( \mathbf{x}\right)
. $

\textbf{Remarks 4.1: }

\begin{enumerate}
\item \emph{The CWF of \cite{Klib97} depends on two large parameters, unlike
the one in (\ref{4.5}). This makes the CWF of \cite{Klib97} to be quite
difficult for calculations, see a similar remark on page 1579 of \cite{Baud}
for the case of a CWF for the hyperbolic operator }$\partial _{t}^{2}-\Delta
.$\emph{\ Thus, we establish in Theorem 4.1 a new Carleman estimate for the
CWF }$e^{2\lambda \left( z+b\right) ^{2}}$\emph{.}

\item \emph{It follows from (\ref{3.180})-(\ref{3.22}) and (\ref{4.4}) that
condition (\ref{4.42}) holds true.}

\item \emph{Analogs of Theorems 4.3-4.6 were not proven in \cite{Klib97}. On
the other hand, Theorems 4.4-4.6 are computationally oriented. }

\item \emph{The regularization parameter }$\beta $\emph{\ was not used in 
\cite{Klib97}. On the other hand, the presence of the regularization term }$%
\beta \left\Vert Q+G\right\Vert _{H^{3}\left( \Omega \right) }^{2}$\emph{\
in the functional }$J_{\lambda ,\beta }\left( Q+G\right) $\emph{\ is
important since this term ensures that in the gradient projection method (%
\ref{4.9}) all functions }$Q_{n}\in H_{0}^{3}\left( \Omega \right) .$ \emph{%
Since }$H^{3}\left( \Omega \right) \subset C^{1}\left( \overline{\Omega }%
\right) ,$\emph{\ and since we use estimates of }$C^{1}\left( \overline{%
\Omega }\right) -$\emph{norms of some functions in the proof of Theorem 4.2,
then we indeed need }$Q_{n}\in H_{0}^{3}\left( \Omega \right) $\emph{.
However, if using the functional of \cite{Klib97}, then there is no
guarantee that }$Q_{n}\in H_{0}^{3}\left( \Omega \right) $.\emph{\ On the
other hand, final estimates (\ref{1})-(\ref{4}) are valid for all values }$%
\beta \in \left( 0,1\right) .$
\end{enumerate}

\section{Proofs}

\label{sec:5}

We now refer to the publication \cite{Bak} where some theorems of convex
analysis are established. The proof of Theorem 4.3 is very similar with the
proof of Theorem 3.1 of \cite{Bak}. As soon as Theorems 4.2 and 4.3 are
proven, the proof of Theorem 4.4 is quite similar with the proof of Lemma
2.1 of \cite{Bak}. Next, as soon as Theorems \ 4.2 and 4.4 are proven, the
proof of Theorem 4.5 is again quite similar with the proof of Theorem 2.1 of 
\cite{Bak}. Hence, we prove here only Theorems 4.1, 4.2 and 4.6.

\subsection{Proof of Theorem 4.1}

\label{sec:5.1}

In this proof, the function $u\in C^{3}\left( \overline{\Omega }\right) .$
The case $u\in H^{2}\left( \Omega \right) $ follows from density arguments.
Consider the function $v=ue^{\lambda \left( z+b\right) ^{2}}.$ Then $%
u=ve^{-\lambda \left( z+b\right) ^{2}}$. Hence, $u_{xx}=v_{xx}e^{-\lambda
\left( z+b\right) ^{2}},u_{yy}=v_{yy}e^{-\lambda \left( z+b\right) ^{2}},$ 
\[
u_{zz}=\left( v_{zz}-2\lambda \left( z+b\right) v_{z}+4\lambda ^{2}\left(
z+b\right) ^{2}\left( 1+O\left( 1/\lambda \right) \right) v\right)
e^{-\lambda \left( z+b\right) ^{2}}. 
\]%
In this proof, $C_{1}=C_{1}\left( \Omega ,b\right) >0$ denotes different
constants depending only on $\Omega $ and $b$ and $O\left( 1/\lambda \right) 
$ denotes different $z-$dependent functions satisfying $\left\vert O\left(
1/\lambda \right) \right\vert ,\left\vert \nabla O\left( 1/\lambda \right)
\right\vert \leq C_{1}/\lambda ,$ Hence,%
\[
\left( \Delta u\right) ^{2}e^{2\lambda \left( z+b\right) ^{2}}=\left[ \left(
v_{xx}+v_{yy}+v_{zz}+4\lambda ^{2}\left( z+b\right) ^{2}\left( 1+O\left(
1/\lambda \right) \right) v\right) -2\lambda \left( z+b\right) v_{z}\right]
^{2} 
\]%
\[
\geq -4\lambda \left( z+b\right) v_{z}\left( v_{xx}+v_{yy}+v_{zz}+4\lambda
^{2}\left( z+b\right) ^{2}\left( 1+O\left( 1/\lambda \right) \right)
v\right) 
\]%
\[
=\left( -4\lambda \left( z+b\right) v_{z}v_{x}\right) _{x}+4\lambda \left(
z+b\right) v_{zx}v_{x}+\left( -2\lambda \left( z+b\right) v_{z}v_{y}\right)
_{y}+4\lambda \left( z+b\right) v_{zy}v_{y} 
\]%
\[
+\left( -2\lambda \left( z+b\right) v_{z}^{2}\right) _{z}+2\lambda
v_{z}^{2}+\left( -8\lambda ^{3}\left( z+b\right) ^{3}\left( 1+O\left(
1/\lambda \right) \right) v^{2}\right) _{z}+24\lambda ^{3}\left( z+b\right)
^{3}\left( 1+O\left( 1/\lambda \right) \right) v^{2} 
\]%
\[
=-2\lambda \left( v_{x}^{2}+v_{y}^{2}\right) +2\lambda v_{z}^{2}+24\lambda
^{3}\left( z+b\right) ^{3}\left( 1+O\left( 1/\lambda \right) \right) v^{2} 
\]%
\[
+\left( -2\lambda \left( z+b\right) v_{z}^{2}+2\lambda \left( z+b\right)
v_{x}^{2}+2\lambda \left( z+b\right) v_{y}^{2}-8\lambda ^{3}\left(
z+b\right) ^{3}\left( 1+O\left( 1/\lambda \right) \right) v^{2}\right) _{z} 
\]%
\[
-2\lambda \left( v_{x}^{2}+v_{y}^{2}\right) +2\lambda v_{z}^{2}+24\lambda
^{3}\left( z+b\right) ^{2}\left( 1+O\left( 1/\lambda \right) \right) v^{2}. 
\]%
Since $v\mid _{\partial \Omega }=v_{z}\mid _{\Gamma _{0}}=0$ and $2\lambda
v_{z}^{2}\geq 0,$ then integrating the above over $\Omega $ going back from $%
v$ to $u$ and using Gauss' formula, we obtain for sufficiently large $%
\lambda \geq C_{1}$%
\begin{equation}
\mathop{\displaystyle \int}\limits_{\Omega }\left( \Delta u\right)
^{2}e^{2\lambda \left( z+b\right) ^{2}}d\mathbf{x}\geq -2\lambda %
\mathop{\displaystyle \int}\limits_{\Omega }\left(
u_{x}^{2}+u_{y}^{2}\right) ^{2}e^{2\lambda \left( z+b\right) ^{2}}d\mathbf{x}%
+23\lambda ^{3}\mathop{\displaystyle \int}\limits_{\Omega }u^{2}e^{2\lambda
\left( z+b\right) ^{2}}d\mathbf{x.}  \label{5.1}
\end{equation}

Next,%
\[
-u\Delta ue^{2\lambda \left( z+b\right) ^{2}}=\left( -u_{x}ue^{2\lambda
\left( z+b\right) ^{2}}\right) _{x}+u_{x}^{2}e^{2\lambda \left( z+b\right)
^{2}}+\left( -u_{y}ue^{2\lambda \left( z+b\right) ^{2}}\right)
_{y}+u_{y}^{2}e^{2\lambda \left( z+b\right) ^{2}} 
\]%
\[
+\left( -u_{z}ue^{2\lambda \left( z+b\right) ^{2}}\right)
_{z}+u_{z}^{2}e^{2\lambda \left( z+b\right) ^{2}}+4\lambda \left( z+b\right)
u_{z}ue^{2\lambda \left( z+b\right) ^{2}} 
\]%
\[
=\left( u_{x}^{2}+u_{y}^{2}+u_{z}^{2}\right) e^{2\lambda \left( z+b\right)
^{2}}+\left( 2\lambda \left( z+b\right) u^{2}e^{2\lambda \left( z+b\right)
^{2}}\right) _{z}-8\lambda ^{2}\left( z+b\right) ^{2}\left( 1+O\left(
1/\lambda \right) \right) u^{2}e^{2\lambda \left( z+b\right) ^{2}} 
\]%
\[
+\left( -u_{x}ue^{2\lambda \left( z+b\right) ^{2}}\right) _{x}+\left(
-u_{y}ue^{2\lambda \left( z+b\right) ^{2}}\right) _{y}. 
\]%
Integrating this over $\Omega $ and using Gauss' formula, we obtain for
sufficiently large $\lambda \geq C_{1}$%
\begin{equation}
-\mathop{\displaystyle \int}\limits_{\Omega }u\Delta ue^{2\lambda \left(
z+b\right) ^{2}}d\mathbf{x}=\mathop{\displaystyle \int}\limits_{\Omega
}\left( u_{x}^{2}+u_{y}^{2}+u_{z}^{2}\right) e^{2\lambda \left( z+b\right)
^{2}}d\mathbf{x}  \label{5.2}
\end{equation}%
\[
-9\lambda ^{2}\mathop{\displaystyle \int}\limits_{\Omega }\left( z+b\right)
^{2}u^{2}e^{2\lambda \left( z+b\right) ^{2}}d\mathbf{x}. 
\]%
Multiply (\ref{5.2}) by $5\lambda /2$ and sum up with (\ref{5.1}). Since $%
23\lambda ^{3}-\left( 9\cdot 5/2\right) \lambda ^{3}=\lambda ^{3}/2,$ then 
\[
\mathbf{-}\frac{5}{2}\lambda \mathop{\displaystyle \int}\limits_{\Omega
}u\Delta ue^{2\lambda \left( z+b\right) ^{2}}d\mathbf{x+}\mathop{%
\displaystyle \int}\limits_{\Omega }\left( \Delta u\right) ^{2}e^{2\lambda
\left( z+b\right) ^{2}}d\mathbf{x} 
\]%
\begin{equation}
\geq \frac{1}{2}\lambda \mathop{\displaystyle \int}\limits_{\Omega }\left(
u_{x}^{2}+u_{y}^{2}+u_{z}^{2}\right) e^{2\lambda \left( z+b\right) ^{2}}d%
\mathbf{x+}\frac{1}{2}\lambda ^{3}\mathop{\displaystyle \int}\limits_{\Omega
}u^{2}e^{2\lambda \left( z+b\right) ^{2}}d\mathbf{x.}  \label{5.3}
\end{equation}%
Next, applying Cauchy-Schwarz inequality, we obtain%
\[
\mathbf{-}\frac{5}{2}\lambda \mathop{\displaystyle \int}\limits_{\Omega
}u\Delta ue^{2\lambda \left( z+b\right) ^{2}}d\mathbf{x+}\mathop{%
\displaystyle \int}\limits_{\Omega }\left( \Delta u\right) ^{2}e^{2\lambda
\left( z+b\right) ^{2}}d\mathbf{x}\leq \frac{25}{4}\lambda ^{2}%
\mathop{\displaystyle \int}\limits_{\Omega }u^{2}e^{2\lambda \left(
z+b\right) ^{2}}d\mathbf{x}+\frac{1}{2}\mathop{\displaystyle \int}%
\limits_{\Omega }\left( \Delta u\right) ^{2}e^{2\lambda \left( z+b\right)
^{2}}d\mathbf{x.} 
\]%
Combining this with (\ref{5.3}), we obtain for sufficiently large $\lambda
_{0}=\lambda _{0}\left( \Omega ,b\right) >0$ and for $\lambda \geq \lambda
_{0}$%
\begin{equation}
\mathop{\displaystyle \int}\limits_{\Omega }\left( \Delta u\right)
^{2}e^{2\lambda \left( z+b\right) ^{2}}d\mathbf{x}\geq C_{1}\lambda %
\mathop{\displaystyle \int}\limits_{\Omega }\left( \left( \nabla u\right)
^{2}+\lambda ^{2}u^{2}\right) e^{2\lambda \left( z+b\right) ^{2}}d\mathbf{x.}
\label{5.4}
\end{equation}

The next step is to incorporate the term with second derivatives in (\ref%
{4.6}). We have 
\begin{equation}
\left( \Delta u\right) ^{2}e^{2\lambda \left( z+b\right) ^{2}}=\left(
u_{xx}+u_{yy}+u_{zz}\right) ^{2}e^{2\lambda \left( z+b\right) ^{2}}=\left(
u_{xx}^{2}+u_{yy}^{2}+u_{zz}^{2}\right) e^{2\lambda \left( z+b\right) ^{2}}
\label{5.5}
\end{equation}%
\[
+2\left( u_{xx}u_{yy}+u_{xx}u_{zz}+u_{yy}u_{zz}\right) e^{2\lambda \left(
z+b\right) ^{2}}. 
\]%
The second line of (\ref{4.6}) gives:%
\[
\left( 2u_{xx}u_{yy}+2u_{xx}u_{zz}+2u_{yy}u_{zz}\right) e^{2\lambda \left(
z+b\right) ^{2}}=\left( 2u_{xx}u_{y}e^{2\lambda \left( z+b\right)
^{2}}\right) _{y}-2u_{xxy}u_{y}e^{2\lambda \left( z+b\right) ^{2}} 
\]%
\[
+\left( 2u_{xx}u_{z}e^{2\lambda \left( z+b\right) ^{2}}\right)
_{z}-2u_{xxz}u_{z}e^{2\lambda \left( z+b\right) ^{2}}-8\lambda \left(
z+b\right) u_{xx}u_{z}e^{2\lambda \left( z+b\right) ^{2}} 
\]%
\begin{equation}
+\left( 2u_{yy}u_{z}e^{2\lambda \left( z+b\right) ^{2}}\right)
_{z}-2u_{yyz}u_{z}e^{2\lambda \left( z+b\right) ^{2}}-8\lambda \left(
z+b\right) u_{yy}u_{z}e^{2\lambda \left( z+b\right) ^{2}}  \label{5.6}
\end{equation}%
\[
=2\left( u_{xy}^{2}+u_{xz}^{2}+u_{yz}^{2}\right) e^{2\lambda \left(
z+b\right) ^{2}}+\left[ 2\left( u_{xy}u_{y}-u_{xz}u_{z}\right) e^{2\lambda
\left( z+b\right) ^{2}}\right] _{x} 
\]%
\[
+\left[ 2\left( u_{xx}u_{y}-u_{xz}u_{z}\right) e^{2\lambda \left( z+b\right)
^{2}}\right] _{y}-8\lambda \left( z+b\right) u_{xx}u_{z}e^{2\lambda \left(
z+b\right) ^{2}}-8\lambda \left( z+b\right) u_{yy}u_{z}e^{2\lambda \left(
z+b\right) ^{2}}. 
\]%
Using Cauchy-Schwarz inequality, we estimate from the below the last line of
(\ref{5.6}) as%
\begin{equation}
-8\lambda \left( z+b\right) u_{xx}u_{z}e^{2\lambda \left( z+b\right)
^{2}}-8\lambda \left( z+b\right) u_{yy}u_{z}e^{2\lambda \left( z+b\right)
^{2}}  \label{5.7}
\end{equation}%
\[
\geq -\frac{1}{2}\left( u_{xx}^{2}+u_{yy}^{2}\right) e^{2\lambda \left(
z+b\right) ^{2}}-64\lambda ^{2}\left( z+b\right) ^{2}u_{z}^{2}e^{2\lambda
\left( z+b\right) ^{2}}. 
\]%
Combining (\ref{5.6})-(\ref{5.7}), we obtain%
\[
\mathop{\displaystyle \int}\limits_{\Omega }\left( \Delta u\right)
^{2}e^{2\lambda \left( z+b\right) ^{2}}d\mathbf{x}\geq \frac{1}{2}%
\mathop{\displaystyle \sum }\limits_{i,j=1}^{3}\mathop{\displaystyle \int}%
\limits_{\Omega }u_{x_{i}x_{j}}^{2}e^{2\lambda \left( z+b\right) ^{2}}d%
\mathbf{x}-64\lambda ^{2}\mathop{\displaystyle \int}\limits_{\Omega }\left(
z+b\right) ^{2}u_{z}^{2}e^{2\lambda \left( z+b\right) ^{2}}d\mathbf{x}. 
\]%
Multiply this estimate by $C_{1}/\left( 128\lambda \right) $ and sum up with
(\ref{5.4}). We obtain%
\begin{equation}
\left( 1+\frac{C_{1}}{128\lambda }\right) \mathop{\displaystyle \int}%
\limits_{\Omega }\left( \Delta u\right) ^{2}e^{2\lambda \left( z+b\right)
^{2}}d\mathbf{x}\geq \frac{C_{1}}{256\lambda }\mathop{\displaystyle \sum }%
\limits_{i,j=1}^{3}\mathop{\displaystyle \int}\limits_{\Omega
}u_{x_{i}x_{j}}^{2}e^{2\lambda \left( z+b\right) ^{2}}d\mathbf{x}
\label{5.8}
\end{equation}%
\[
\mathbf{+}\frac{C}{2}\lambda \mathop{\displaystyle \int}\limits_{\Omega
}\left( \left( \nabla u\right) ^{2}+\lambda ^{2}u^{2}\right) e^{2\lambda
\left( z+b\right) ^{2}}d\mathbf{x.} 
\]%
Since $C_{1}>0$ denotes different constants, then the target estimate (\ref%
{4.6}) follows from (\ref{4.8}) immediately. $\square $

\subsection{Proof of Theorem 4.2}

\label{sec:5.2}

Denote $h=Q_{2}-Q_{1}$ implying that $Q_{2}=Q_{1}+h.$ Also, $h\in
H_{0}^{3}\left( \Omega \right) ,$ $\left\Vert h\right\Vert _{H^{3}\left(
\Omega \right) }<2R$.\ Hence, by (\ref{4.40}) 
\begin{equation}
\left\Vert h\right\Vert _{C^{1}\left( \overline{\Omega }\right)
}<D_{1}\left( 2R\right) .  \label{5.9}
\end{equation}%
Using the multidimensional analog of Taylor formula (see, e.g. \cite{V} for
this formula)\emph{\ }and (\ref{4.42}), we obtain 
\[
\Delta h+\left( \Delta Q_{1}+\Delta G\right) +F\left( h+Q_{1}+G,\nabla
\left( h+Q_{1}+G\right) \right) 
\]%
\begin{equation}
=\left[ \Delta h+F^{\left( 1\right) }\left( Q_{1}+G,\nabla \left(
Q_{1}+G\right) \right) h+F^{\left( 2\right) }\left( Q_{1}+G,\nabla \left(
Q_{1}+G\right) \right) \nabla h\right]  \label{5.10}
\end{equation}%
\[
+F_{nonlin}\left( h,\nabla h,Q_{1}+G,\nabla \left( Q_{1}+G\right) \right) + 
\left[ \left( \Delta Q_{1}+\Delta G\right) +F\left( Q_{1}+G,\nabla \left(
Q_{1}+G\right) \right) \right] , 
\]%
where elements of $\left( N+1\right) \times \left( N+1\right) $ matrix $%
F^{\left( 1\right) }$ and $\left( 3N+3\right) \times \left( 3N+3\right) $
matrix are bounded for $\mathbf{x}\in \overline{\Omega }$, i.e.%
\begin{equation}
\left\vert F_{i,j}^{\left( 1\right) }\left( Q_{1}+G,\nabla \left(
Q_{1}+G\right) \right) \right\vert ,\left\vert F_{i,j}^{\left( 2\right)
}\left( Q_{1}+G,\nabla \left( Q_{1}+G\right) \right) \right\vert \leq C_{2},%
\text{ }\forall \mathbf{x}\in \overline{\Omega },  \label{5.11}
\end{equation}%
where the subscript \textquotedblleft $i,j"$ denotes an arbitrary element of
the corresponding matrix indexed as $\left( i,j\right) .$ Next, the $\left(
N+1\right) -$dimensional vector function $F_{nonlin}$ depends nonlinearly on 
$h,\nabla h.$ Furthermore, the following estimate follows from (\ref{4.40})-(%
\ref{4.42})%
\begin{equation}
\left\vert F_{nonlin}\left( h,\nabla h,Q_{1}+G,\nabla \left( Q_{1}+G\right)
\right) \right\vert \leq C_{2}\left( \left\vert h\right\vert ^{2}+\left\vert
\nabla h\right\vert ^{2}\right) ,\text{ }\forall \mathbf{x}\in \overline{%
\Omega }.  \label{5.12}
\end{equation}%
Next, (\ref{5.9}) and (\ref{5.12}) imply with a different constant $C_{2}$ 
\begin{equation}
\left\vert F_{nonlin}\left( h,\nabla h,Q_{1}+G,\nabla \left( Q_{1}+G\right)
\right) \right\vert \leq C_{2}\left( \left\vert h\right\vert +\left\vert
\nabla h\right\vert \right) ,\text{ }\forall \mathbf{x}\in \overline{\Omega }%
.  \label{5.13}
\end{equation}%
It follows from (\ref{5.10})-(\ref{5.13}) that%
\[
\left[ \Delta h+\left( \Delta Q_{1}+\Delta G\right) +F\left(
h+Q_{1}+G,\nabla \left( h+Q_{1}+G\right) \right) \right] ^{2} 
\]%
\begin{equation}
-\left[ \left( \Delta Q_{1}+\Delta G\right) +F\left( Q_{1}+G,\nabla \left(
Q_{1}+G\right) \right) \right] ^{2}  \label{5.14}
\end{equation}%
\[
=Lin_{1}\left( \Delta h\right) +Lin_{2}\left( \nabla h\right) +Lin_{3}\left(
h\right) 
\]%
\[
+\left( \Delta h\right) ^{2}+M_{1}\left( h,\nabla h,Q_{1}+G,\nabla \left(
Q_{1}+G\right) \right) \Delta h+M_{2}\left( h,\nabla h,Q_{1}+G,\nabla \left(
Q_{1}+G\right) \right) , 
\]%
where expressions $Lin_{1}\left( \Delta h\right) ,Lin_{2}\left( \nabla
h\right) $ and $Lin_{3}\left( h\right) $ are linear with respect to $\Delta
h,\nabla h$ and $h$ respectively and 
\begin{equation}
\left\vert Lin_{1}\left( \Delta h\right) +Lin_{2}\left( \nabla h\right)
+Lin_{3}\left( h\right) \right\vert \leq C_{2}\left( \left\vert \Delta
h\right\vert +\left\vert \nabla h\right\vert +\left\vert h\right\vert
\right) ,\text{ }\forall \mathbf{x}\in \overline{\Omega }.  \label{5.15}
\end{equation}%
Next, the following estimates are valid for $M_{1}$ and $M_{2}$ 
\begin{equation}
\left\vert M_{1}\left( h,\nabla h,Q_{1}+G,\nabla \left( Q_{1}+G\right)
\right) \right\vert \leq C_{2}\left( \left\vert \nabla h\right\vert
+\left\vert h\right\vert \right) ,\text{ }\forall \mathbf{x}\in \overline{%
\Omega },  \label{5.16}
\end{equation}%
\begin{equation}
\left\vert M_{2}\left( h,\nabla h,Q_{1}+G,\nabla \left( Q_{1}+G\right)
\right) \right\vert \leq C_{2}\left( \left\vert \nabla h\right\vert
^{2}+\left\vert h\right\vert ^{2}\right) ,\forall \mathbf{x}\in \overline{%
\Omega }.  \label{5.17}
\end{equation}%
In particular, (\ref{5.16}), (\ref{5.17}) and Cauchy-Schwarz inequality
imply 
\[
\left( \Delta h\right) ^{2}+M_{1}\left( h,\nabla h,Q_{1}+G,\nabla \left(
Q_{1}+G\right) \right) \Delta h+M_{2}\left( h,\nabla h,Q_{1}+G,\nabla \left(
Q_{1}+G\right) \right) 
\]%
\begin{equation}
\geq \frac{1}{2}\left( \Delta h\right) ^{2}-C_{2}\left( \left\vert \nabla
h\right\vert ^{2}+\left\vert h\right\vert ^{2}\right) ,\forall \mathbf{x}\in 
\overline{\Omega }.  \label{5.18}
\end{equation}%
Using (\ref{4.5}) and (\ref{5.14})-(\ref{5.17}), we obtain%
\begin{equation}
J_{\lambda ,\beta }\left( Q_{1}+h\right) -J_{\lambda ,\beta }\left(
Q_{1}\right) =X_{lin}\left( h\right) +X_{nonlin}\left( h\right) ,
\label{5.19}
\end{equation}%
where $X_{lin}\left( h\right) $ can be extended from $\left\{ \left\Vert
h\right\Vert _{H^{3}\left( \Omega \right) }<2R\right\} \subset
H_{0}^{3}\left( \Omega \right) $ to the entire space $H^{3}\left( \Omega
\right) $ as a bounded linear \ functional, 
\begin{equation}
X_{lin}\left( h\right) =e^{-2\lambda b^{2}}\mathop{\displaystyle \int}%
\limits_{\Omega }\left( Lin_{1}\left( \Delta h\right) +Lin_{2}\left( \nabla
h\right) +Lin_{3}\left( h\right) \right) \left( \mathbf{x}\right)
e^{2\lambda \left( z+b\right) ^{2}}d\mathbf{x}+2\beta \left[ h,Q_{1}+G\right]
,  \label{5.20}
\end{equation}%
where $\left[ .,.\right] $ is the scalar product in $H^{3}\left( \Omega
\right) .$ As to $X_{nonlin}\left( h\right) $ in (\ref{5.19}), it follows
from (\ref{4.5}), (\ref{5.14}), (\ref{5.16}) and (\ref{5.17}) that 
\begin{equation}
\lim_{\left\Vert h\right\Vert _{H^{3}\left( \Omega \right) }\rightarrow 0}%
\frac{X_{nonlin}\left( h\right) }{\left\Vert h\right\Vert _{H^{3}\left(
\Omega \right) }}=0.  \label{5.21}
\end{equation}%
Using (\ref{5.15}) and (\ref{5.19})-(\ref{5.21}), we obtain that $%
X_{lin}\left( h\right) $ is the Frech\'{e}t derivative $J_{\lambda ,\beta
}^{\prime }\left( Q\right) $ of the functional $J_{\lambda ,\beta }\left(
Q\right) $ at the point $Q$, i.e. $X_{lin}\left( h\right) =J_{\lambda ,\beta
}^{\prime }\left( Q_{1}\right) \left( h\right) $. Thus, the existence of the
Frech\'{e}t derivative is established.

Next, using (\ref{4.5}) and (\ref{5.14})-(\ref{5.20}), we obtain%
\begin{equation}
J_{\lambda ,\beta }\left( Q_{1}+G+h\right) -J_{\lambda ,\beta }\left(
Q_{1}+G\right) -J_{\lambda ,\beta }^{\prime }\left( Q_{1}+G\right) \left(
h\right)  \label{5.22}
\end{equation}%
\[
\geq \frac{1}{2}e^{-2\lambda b^{2}}\mathop{\displaystyle \int}%
\limits_{\Omega }\left( \Delta h\right) ^{2}e^{2\lambda \left( z+b\right)
^{2}}d\mathbf{x}-C_{2}e^{-2\lambda b^{2}}\mathop{\displaystyle \int}%
\limits_{\Omega }\left( \left\vert \nabla h\right\vert ^{2}+\left\vert
h\right\vert ^{2}\right) e^{2\lambda \left( z+b\right) ^{2}}d\mathbf{x}%
+\beta \left\Vert h\right\Vert _{H^{3}\left( \Omega \right) }^{2}. 
\]%
We now apply Carleman estimate (\ref{4.6}), assuming that $\lambda \geq
\lambda _{0},$%
\[
\frac{1}{2}e^{-2\lambda b^{2}}\mathop{\displaystyle \int}\limits_{\Omega
}\left( \Delta h\right) ^{2}e^{2\lambda \left( z+b\right) ^{2}}d\mathbf{x}%
-C_{2}e^{-2\lambda b^{2}}\mathop{\displaystyle \int}\limits_{\Omega }\left(
\left\vert \nabla h\right\vert ^{2}+\left\vert h\right\vert ^{2}\right)
e^{2\lambda \left( z+b\right) ^{2}}d\mathbf{x}+\beta \left\Vert h\right\Vert
_{H^{3}\left( \Omega \right) }^{2} 
\]%
\[
\geq \frac{C_{1}}{\lambda }\mathop{\displaystyle \sum }%
\limits_{i,j=1}^{3}e^{-2\lambda b^{2}}\mathop{\displaystyle \int}%
\limits_{\Omega }h_{x_{i}x_{j}}^{2}e^{2\lambda \left( z+b\right) ^{2}}d%
\mathbf{x+}C_{1}\lambda e^{-2\lambda b^{2}}\mathop{\displaystyle \int}%
\limits_{\Omega }\left( \left( \nabla h\right) ^{2}+\lambda ^{2}h^{2}\right)
e^{2\lambda \left( z+b\right) ^{2}}d\mathbf{x} 
\]%
\[
-C_{2}e^{-2\lambda b^{2}}\mathop{\displaystyle \int}\limits_{\Omega }\left(
\left\vert \nabla h\right\vert ^{2}+\left\vert h\right\vert ^{2}\right)
e^{2\lambda \left( z+b\right) ^{2}}d\mathbf{x}+\beta \left\Vert h\right\Vert
_{H^{3}\left( \Omega \right) }^{2}. 
\]%
Choosing sufficiently large $\lambda _{1}=\lambda _{1}\left( F,\left\Vert
G\right\Vert _{H^{3}\left( \Omega \right) },m,R,\Omega ,b\right) \geq
\lambda _{0}$ and letting $\lambda \geq \lambda _{1},$ we obtain with a
different constant $C_{2}$%
\[
\frac{1}{2}e^{-2\lambda b^{2}}\mathop{\displaystyle \int}\limits_{\Omega
}\left( \Delta h\right) ^{2}e^{2\lambda \left( z+b\right) ^{2}}d\mathbf{x}%
-C_{2}e^{-2\lambda b^{2}}\mathop{\displaystyle \int}\limits_{\Omega }\left(
\left\vert \nabla h\right\vert ^{2}+\left\vert h\right\vert ^{2}\right)
e^{2\lambda \left( z+b\right) ^{2}}d\mathbf{x}+\beta \left\Vert h\right\Vert
_{H^{3}\left( \Omega \right) }^{2} 
\]%
\[
\geq \frac{C_{2}}{\lambda }\mathop{\displaystyle \sum }%
\limits_{i,j=1}^{3}e^{-2\lambda b^{2}}\mathop{\displaystyle \int}%
\limits_{\Omega }h_{x_{i}x_{j}}^{2}e^{2\lambda \left( z+b\right) ^{2}}d%
\mathbf{x+}C_{2}\lambda e^{-2\lambda b^{2}}\mathop{\displaystyle \int}%
\limits_{\Omega }\left( \left( \nabla h\right) ^{2}+\lambda ^{2}h^{2}\right)
e^{2\lambda \left( z+b\right) ^{2}}d\mathbf{x}+\beta \left\Vert h\right\Vert
_{H^{3}\left( \Omega \right) }^{2}. 
\]%
This, (\ref{5.22}) and (\ref{4.50}) imply (\ref{4.7}). \ $\square $

\subsection{Proof of Theorem 4.6}

\label{sec:5.3}

We rewrite the functional $J_{\lambda ,\beta }\left( Q\right) $ in (\ref{4.5}%
) as 
\begin{equation}
J_{\lambda ,\beta }\left( Q+G\right) =J_{\lambda ,\beta }^{0}\left(
Q+G\right) +\beta \left\Vert Q+G\right\Vert _{H^{3}\left( \Omega \right)
}^{2}.  \label{5.23}
\end{equation}%
Since the vector function $Q^{\ast }\in B\left( m,R\right) $ is the exact
solution of the problem (\ref{4.2}), (\ref{4.3}) with the noiseless data $%
G^{\ast },$ then $J_{\lambda ,\beta }^{0}\left( Q^{\ast }+G^{\ast }\right)
=0.$ Hence, 
\begin{equation}
J_{\lambda ,\beta }\left( Q^{\ast }+G^{\ast }\right) \leq C_{2}\beta .
\label{5.24}
\end{equation}%
Next, $J_{\lambda ,\beta }\left( Q^{\ast }+G\right) =\left( J_{\lambda
,\beta }\left( Q^{\ast }+G\right) -J_{\lambda ,\beta }\left( Q^{\ast
}+G^{\ast }\right) \right) +J_{\lambda ,\beta }\left( Q^{\ast }+G^{\ast
}\right) .$ Hence, applying (\ref{5.24}), we obtain%
\begin{equation}
J_{\lambda ,\beta }\left( Q^{\ast }+G\right) \leq \left\vert J_{\lambda
,\beta }\left( Q^{\ast }+G\right) -J_{\lambda ,\beta }\left( Q^{\ast
}+G^{\ast }\right) \right\vert +C_{2}\beta .  \label{5.25}
\end{equation}%
Using (\ref{4.10}) and (\ref{5.23}), we estimate the first term in the right
hand side of (\ref{5.25}),%
\[
\left\vert J_{\lambda ,\beta }\left( Q^{\ast }+G\right) -J_{\lambda ,\beta
}\left( Q^{\ast }+G^{\ast }\right) \right\vert \leq \left\vert J_{\lambda
,\beta }^{0}\left( Q^{\ast }+G\right) -J_{\lambda ,\beta }^{0}\left( Q^{\ast
}+G^{\ast }\right) \right\vert +C_{2}\beta \delta 
\]%
\begin{equation}
\leq C_{2}\delta \exp \left( 2\lambda \left( A+b\right) ^{2}\right)
+C_{2}\beta \delta .  \label{5.26}
\end{equation}%
Recall that due to our choice $\lambda =\lambda \left( \delta \right) =\ln
\delta ^{-\eta },$ where $\eta =\left[ 4\left( A+b\right) ^{2}\right] ^{-1},$
we have $\delta \exp \left( 2\lambda \left( A+b\right) ^{2}\right) =1/\sqrt{%
\delta }.$ Hence, (\ref{5.26}) implies 
\[
\left\vert J_{\lambda ,\beta }\left( Q^{\ast }+G\right) -J_{\lambda ,\beta
}\left( Q^{\ast }+G^{\ast }\right) \right\vert \leq C_{2}\sqrt{\delta },%
\text{ }\forall \beta \in \left( 0,1\right) . 
\]%
Combining this with (\ref{5.24}), we obtain%
\begin{equation}
J_{\lambda ,\beta }\left( Q^{\ast }+G\right) \leq C_{2}\left( \sqrt{\delta }%
+\beta \right) .  \label{5.27}
\end{equation}

Until now we have not used in this proof the strict convexity of the
functional $J_{\lambda ,\beta }\left( Q+G\right) $ for $Q\in B\left(
m,R\right) .$ But we will use this property in the rest of the proof. Recall
that by Theorem 4.3 $Q_{\min ,\lambda ,\beta }\in \overline{B\left(
m,R\right) }$ is the unique minimizer on the set $\overline{B\left(
m,R\right) }$ of the functional $J_{\lambda ,\beta }\left( Q+G\right) $ on
the set $\overline{B\left( m,R\right) }$. By Theorem 4.2 
\begin{equation}
J_{\lambda ,\beta }\left( Q^{\ast }+G\right) -J_{\lambda ,\beta }\left(
Q_{\min ,\lambda ,\beta }+G\right) -J_{\lambda ,\beta }^{\prime }\left(
Q_{\min ,\lambda ,\beta }\right) \left( Q^{\ast }-Q_{\min ,\lambda ,\beta
}\right)  \label{5.28}
\end{equation}%
\[
\geq \frac{C_{2}}{\lambda }\mathop{\displaystyle \sum }\limits_{i,j=1}\left%
\Vert \left( Q^{\ast }-Q_{\min ,\lambda ,\beta }\right)
_{x_{i}x_{j}}\right\Vert _{L_{2}\left( \Omega \right) }^{2}+C_{2}\lambda
\left\Vert Q^{\ast }-Q_{\min ,\lambda ,\beta }\right\Vert _{H^{1}\left(
\Omega \right) }^{2}+\frac{\beta }{2}\left\Vert Q^{\ast }-Q_{\min ,\lambda
,\beta }\right\Vert _{H^{3}\left( \Omega \right) }^{2}. 
\]%
Since by (\ref{4.8}) $-J_{\lambda ,\beta }^{\prime }\left( Q_{\min ,\lambda
,\beta }\right) \left( Q^{\ast }-Q_{\min ,\lambda ,\beta }\right) \leq 0,$
then (\ref{5.27}) implies that the left hand side of (\ref{5.28}) can be
estimated as%
\[
J_{\lambda ,\beta }\left( Q^{\ast }+G\right) -J_{\lambda ,\beta }\left(
Q_{\min ,\lambda ,\beta }+G\right) -J_{\lambda ,\beta }^{\prime }\left(
Q_{\min ,\lambda ,\beta }+G\right) \left( Q^{\ast }-Q_{\min ,\lambda ,\beta
}\right) \leq C_{2}\left( \sqrt{\delta }+\beta \right) . 
\]%
Hence, using our choice of $\lambda =\lambda \left( \delta \right) =\ln
\delta ^{-\eta }$ and (\ref{5.28}), we obtain estimates (\ref{1}) and (\ref%
{2}). Estimates (\ref{3}) and (\ref{4}) are obtained from (\ref{1}) and (\ref%
{2}) respectively using (\ref{4.90}) and the triangle inequality. Estimate (%
\ref{5}) obviously follows from estimate (\ref{4}). $\ \square $

\section{Numerical Studies}

\label{sec:6}

The single point source is now $\mathbf{x}_{0}=(0,0,-5).$ We choose in (\ref%
{2.1}) the numbers $A=1$. Hence, below 
\[
\Omega =\left\{ -1/2<x,y<1/2,z\in \left( 0,1\right) \right\} , 
\]%
\begin{equation}
\Gamma _{0}=\left\{ \mathbf{x=}\left( x,y,z\right) :-1/2<x,y<1/2,z=1\right\}
,\Gamma _{1}=\partial \Omega \diagdown \Gamma _{0}.  \label{6.0}
\end{equation}%
We have introduced the vector function $G$ in section 4.1 and thus obtained
the problem (\ref{4.2}), (\ref{4.3}) for the vector function $Q=W-G$ from
the problem (\ref{3.22}), (\ref{3.23}) for the vector function $W$ in order
to obtain zero boundary conditions (\ref{4.3}) for $Q$. The latter was
convenient for proofs of Theorems 4.3-4.6. However, it follows from Theorems
4.2-4.6 that their obvious analogs hold true for the functional 
\begin{equation}
J_{\lambda ,\beta }\left( W\right) =e^{-2\lambda b^{2}}\mathop{\displaystyle
\int}\limits_{\Omega }\left( \Delta W+F\left( \nabla W,W\right) \right)
^{2}e^{2\lambda \left( z+b\right) ^{2}}d\mathbf{x}+\beta \left\Vert
W\right\Vert _{H^{3}\left( \Omega \right) }^{2},  \label{6.1}
\end{equation}%
\begin{equation}
W\in B_{W}\left( m,R\right) =\left\{ W:W=Q-G,\forall Q\in B\left( m,R\right)
\right\} .  \label{6.10}
\end{equation}%
Furthermore, we use in (\ref{6.1}) $\beta =0,b=0$. Therefore, we ignore the
multiplier $e^{-2\lambda b^{2}},$ which was used above to balance first and
second terms in the right hand side of (\ref{4.5}), see (\ref{4.50}). Hence,
we minimize the weighted cost functional 
\begin{equation}
J_{\lambda }\left( W\right) =\mathop{\displaystyle \int}\limits_{\Omega
}\left( \Delta W+F\left( \nabla W,W\right) \right) ^{2}e^{2\lambda z^{2}}d%
\mathbf{x}  \label{6.100}
\end{equation}%
on the set (\ref{6.10}). We conjecture that the case $\beta =0$ works
probably because the minimal mesh size of 1/32 in the finite differences we
use to minimize this functional is not too small, and all norms in finite
dimensional spaces are equivalent, see item 4 of Remark 4.1. In addition,
recall that, by the same item, one can choose any value of $\beta \in \left(
0,1\right) $ in convergence estimates (\ref{1})-(\ref{4}). Also, we use the
gradient descent method (GD) instead of a more complicated gradient
projection method.\ We have observed that GD works well for our
computations. The latter coincides with observations in all earlier
publications about numerical studies of the convexification \cite%
{KlibThanh,KlibKol1,KlibKol2,KlibKol3,KlibKol4,KEIT}. As to our choice $b=0,$
one can derive from the proof of Theorem 4.1 that a slightly modified
Carleman estimate of this theorem works in a little bit smaller domain $%
\Omega ^{\prime }=\Omega \cap \left\{ z>\varepsilon \right\} $ for any small
number $\varepsilon >0.$ Finally, we believe that simplifications listed in
this section work well numerically due to a commonly known observation that
numerical studies are usually less pessimistic than the theory is.

\subsection{Some details of the numerical implementation}

\label{sec:6.1}

To solve the inverse problem, we should first computationally simulate the
data (\ref{2.10}) at $\partial \Omega $ via the numerical solution of the
forward problem (\ref{2.7}), (\ref{2.8}). To solve the problem (\ref{2.7}), (%
\ref{2.8}) computationally, we have used the standard finite difference
method. To avoid the use of the infinite space $\mathbb{R}^{3}$ in the
solution of the forward problem, we choose the cube $\Omega _{f}=\left\{
-6.5<x,y<6.5,z\in \left( -6,7\right) \right\} .$ So that $\Omega \subset
\Omega _{f}$, $\partial \Omega \cap \partial \Omega _{f}=\varnothing $ and $%
\mathbf{x}_{0}=\left( 0,0,-5\right) \in \Omega _{f}.$ We choose a
sufficiently large number $T_{0}=6.5$. Then we solve equation (\ref{2.7})
with the initial condition (\ref{2.8}) and zero Dirichlet boundary condition
at $\partial \Omega _{f}$ for $\left( \mathbf{x},t\right) \in \Omega
_{f}\times \left( 0,T_{0}\right) $ via finite differences. Indeed, the wave
originated at $\mathbf{x}_{0}$ cannot reach neither vertical sides of $%
\Omega _{f}$ nor the upper side $\left\{ z=7\right\} \cap \overline{\Omega
_{f}}$ of $\Omega $ for times $t\in \left( 0,6.5\right) .$ However, it
reaches the upper side $\left\{ z=1\right\} \cap \overline{\Omega }$ of $%
\Omega $ at $t=6.$ This wave reaches the lower side $\left\{ z=-6\right\}
\cap \overline{\Omega _{f}}$ \ of $\Omega _{f}$ at $t=1,$ which means that
the zero Dirichlet boundary condition on the lower side is incorrect. Still,
for $t\in \left( 0,6.5\right) ,$ the wave reflected from the lower side of
of $\Omega _{f}$ does not reach the upper side $\left\{ z=1\right\} \cap 
\overline{\Omega }$ of $\Omega ,$ where the \ data for our CIP are given.
Hence, this reflected wave does not affect our data.

We use the explicit scheme. The grid step size in each spatial direction is $%
\Delta x=1/32$ and in time direction $\Delta t=0.002.$ To avoid a
substantial increase of the computational time, we do not decrease these
step sizes. When solving the forward problem, we model the $\delta \left( 
\mathbf{x}-\mathbf{x}_{0}\right) -$function in (\ref{2.8}) as%
\[
\widetilde{\delta }\left( \mathbf{x}-\mathbf{x}_{0}\right) =\left\{ 
\begin{array}{l}
\frac{1}{\varepsilon }\exp \left( -\frac{1}{1-\left\vert \mathbf{x}-\mathbf{x%
}_{0}\right\vert ^{2}/\varepsilon }\right) ,\text{ if }\left\vert \mathbf{x}-%
\mathbf{x}_{0}\right\vert ^{2}<\varepsilon =0.01, \\ 
0,\text{ otherwise.}%
\end{array}%
\right. 
\]

%
%

In computations of the inverse problem, for each test we use, we choose in
the data (\ref{2.10}) $T=\max_{\overline{\Omega }}\tau \left( \mathbf{x}%
\right) +0.1.$ We have observed that $T<T_{0}$ in all our tests. Next, we
set $T_{1}=T-\max_{\overline{\Omega }}\tau \left( \mathbf{x}\right) =0.1,$
see (\ref{3.11}), (\ref{3.110}). An important question is on how do we
figure out boundary conditions at $\partial \Omega $ for the function $\tau
\left( \mathbf{x}\right) ,$ i.e. $\tau \left( \mathbf{x}\right) \mid
_{\partial \Omega }$ and also $\partial _{z}\tau \left( \mathbf{x}\right)
\mid _{\Gamma _{0}}.$ In principle, for $\mathbf{x}\in \partial \Omega ,$
one should choose such a number $\tau _{0}\left( \mathbf{x}\right) $ that $%
\tau _{0}\left( \mathbf{x}\right) =\min_{t}\left\{ t:u\left( \mathbf{x,}%
t\right) >0\right\} .$ However, it is hard to choose in practice the number $%
\tau _{0}\left( \mathbf{x}\right) $ satisfying this criterion. Therefore, we
calculate such a number $\widetilde{t}\left( \mathbf{x}\right) $ at which
the first wave with the largest maximal value arrives at the point $\mathbf{x%
}\in \partial \Omega $, see Figure 2. Next, we set $\tau _{0}\left( \mathbf{x%
}\right) \mid _{\partial \Omega }:=\widetilde{t}\left( \mathbf{x}\right)
\mid _{\partial \Omega }.$ To calculate the derivative $\partial _{z}\tau
_{0}\left( \mathbf{x}\right) \mid _{\Gamma _{0}},$ we compute the discrete
normal derivative of $\tau _{0}\left( \mathbf{x}\right) $ over the mesh in
the forward problem.

\begin{figure}[tbp]
\begin{center}
\begin{tabular}{cc}
\includegraphics[width=6cm]{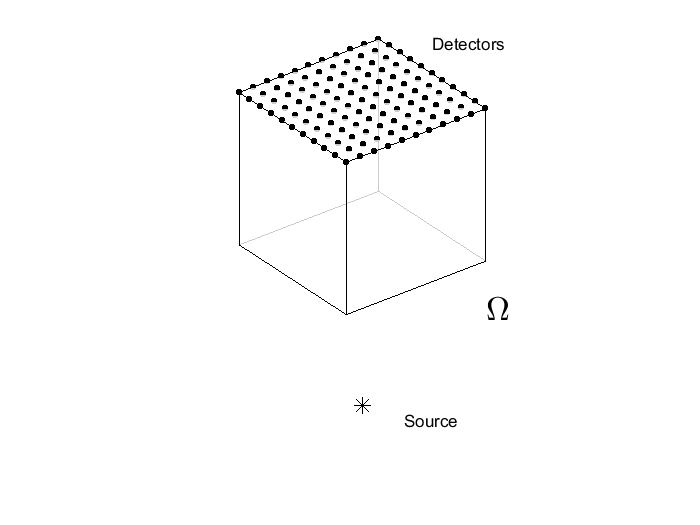} & %
\includegraphics[width=6cm]{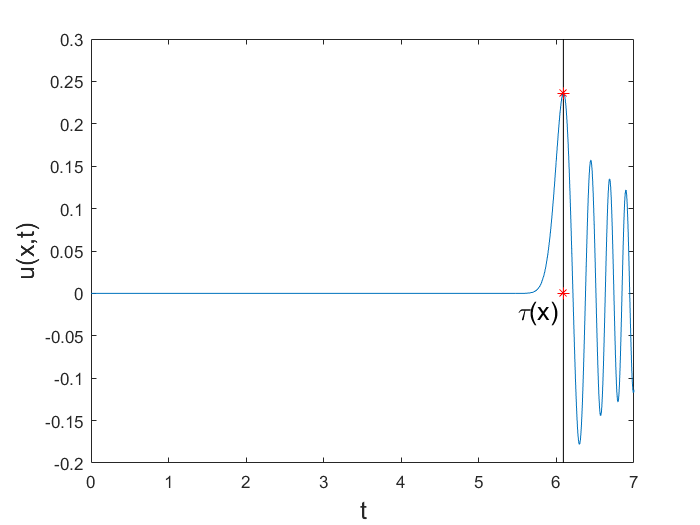} \\ 
(a) & (b)%
\end{tabular}%
\end{center}
\caption{(a) A schematic diagram of domains $\Omega ,$ source and detectors.
(b) This figure explains how do we approximately choose the boundary
condition $\protect\tau \left( \mathbf{x}\right) \mid _{\partial \Omega }.$
We have chosen here a selected point ${\mathbf{x}}\in \Gamma _{0}$}
\label{domain}
\end{figure}

To minimize the weighted cost functional $J_{\lambda }\left( W\right) $ in (%
\ref{6.10}), we act similarly with the previous above cited works about
numerical studies of the convexification for a number of other CIPs. More
precisely, we write the differential operators involved in $J_{\lambda
}\left( W\right) $ via finite differences and minimize with respect to the
values of the discrete analog of the vector function $W$ at grid points. As
to the choice of the parameter $\lambda ,$even though the above theory works
only for sufficiently large values of $\lambda $, we have established in our
computational experiments that the choice $\lambda =1$ is an optimal one for
all tests we have performed. This again repeats observation of all above
cited works on numerical studies of the convexification, in which the
optimal choice was $\lambda \in \left[ 1,3\right] .$ We have also tested two
different values of the number $N$ terms in the series (\ref{3.16}): $N=1$
and $N=3.$ Our computational results indicate that $N=3$ provides results of
a good quality. In all tests below, the starting point of GD is the vector
function $W\left( \mathbf{x}\right) ,$ which is generated by the coefficient 
$c\left( \mathbf{x}\right) \equiv 1$ in equation (\ref{2.7}), $W_{c\equiv
1}\left( \mathbf{x}\right) .$

\subsection{A multi-level minimization method of the functional $J_{\protect%
\lambda }\left( W\right) $}

\label{sec:6.2}

We have found in our computational experiments that the gradient descent
method for our weighted cost functional $J_{\lambda ,0}\left( W\right) $
converges rapidly on a coarse mesh. This provides us with a rough image.
Hence, we have implemented a multi-level method \cite{LZ}. Let $%
M_{h_{1}}\subset M_{h_{2}}...\subset M_{h_{K}}$ be nested finite difference
meshes, i.e. $M_{h_{k}}$ is a refinement of $M_{h_{k-1}}$ for $k\leq K$. Let 
$P_{h_{k}}$ be the corresponding finite difference functional space. One the
first level $M_{h_{1}}$, we solve the discrete optimization problem. In
other words, let $W_{h_{1},\min }$ be the minimizer on the finite difference
analog of the set (\ref{6.10}) of the following functional 
\begin{equation}
J_{\lambda }\left( W_{h_{1}}\right) =\mathop{\displaystyle \int}%
\limits_{\Omega }\left( \Delta W_{h_{1}}+F\left( \nabla
W_{h_{1}},W_{h_{1}}\right) \right) ^{2}e^{2\lambda z^{2}}d\mathbf{x.}
\label{6.4}
\end{equation}%
In (\ref{6.4}) the integral and the derivatives are understood in the
discrete sense, and $W_{h_{1},\min }$ is found via the GD. Then we
interpolate the minimizer $W_{h_{1},\min }$ on the finer mesh $M_{h_{2}}$
and use the resulting vector function $W_{h_{2},\text{int}}$ as the starting
point of the gradient descent method of the optimization of the direct
analog of functional (\ref{6.4}) in which $h_{1}$ is replaced with $h_{2}$
and $W_{h_{1}}$ is replaced with $W_{h_{2}}.$ This process is repeated until
we obtain the minimizer $W_{h_{K},\min }$ on the $K_{th}$ level on the mesh $%
M_{h_{K}}$.

Since $(x,y,z)\in (-1/2,1/2)\times (-1/2,1/2)\times (0,1)$, then our first
level $M_{h_{1}}$ is set to be the uniform mesh with the grid step $%
h_{1}=1/8 $. For each mesh refinement, we will refine the mesh via setting
the new grid step of the refined mesh in all directions to be 1/2 of the
previous grid step. On each level $M_{h_{k}}$, we stop iterations as soon as
we see that $\Vert \nabla J_{\lambda }^{(h_{k})}(W_{h_{k}})\Vert <2\times
10^{-2}$. Next, we refine the mesh and compute the solution on the next
level $M_{h_{k+1}}$. In the end, we compute our approximation for the target
coefficient $c(\mathbf{x})$ using the final minimizer $W_{h_{K},\min }$.

\subsection{Numerical testing}

\label{sec:6.3}

In the tests of this section, we demonstrate the efficiency of our numerical
method for imaging of small inclusions as well as for imaging of a smoothly
varying function $c\left( \mathbf{x}\right) $. In all tests the background
value of $c_{bkgr}=1.$ Note that a postprocessing of images was not applied
in our numerical tests. All necessary derivatives of the data were
calculated using finite differences, just as in all above cited previous
publications about numerical studies of the convexification, including two
noisy experimental data \cite{KlibKol3,KlibKol4}. Just as in all those
works, we have not observed instabilities due to the differentiation, most
likely because the step sizes of finite differences were not too small. On
Figures 2-7 slices are depicted to demonstrate the values of the computed
function $c\left( \mathbf{x}\right) .$

\begin{figure}[tbp]
\begin{center}
\begin{tabular}{cc}
\includegraphics[width=5cm]{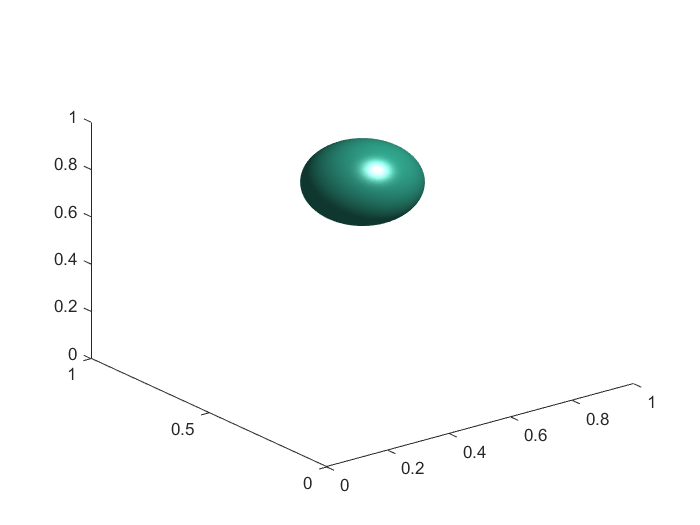} & %
\includegraphics[width=5cm]{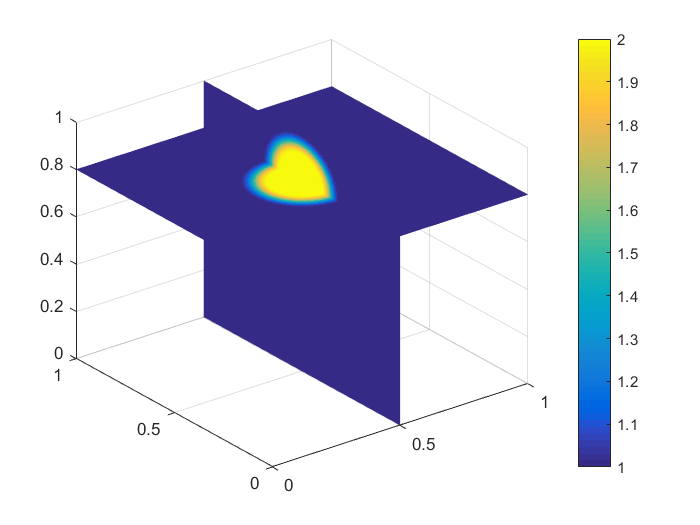} \\ 
(a)3D image of true $c$ & (b) Slice image of true $c$ \\ 
\includegraphics[width=5cm]{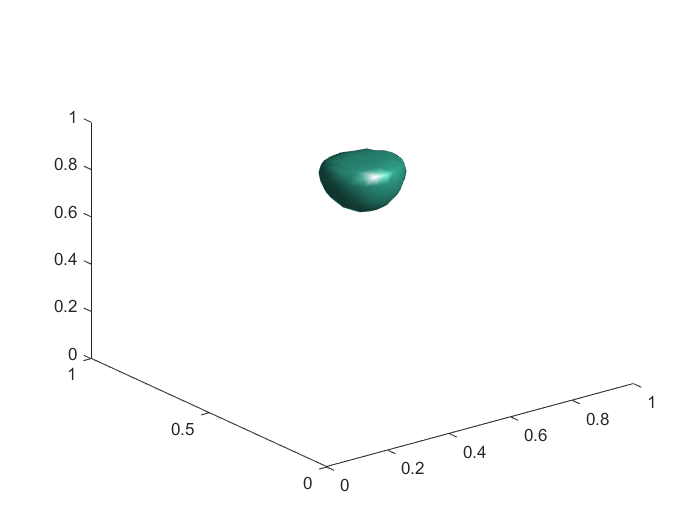} & %
\includegraphics[width=5cm]{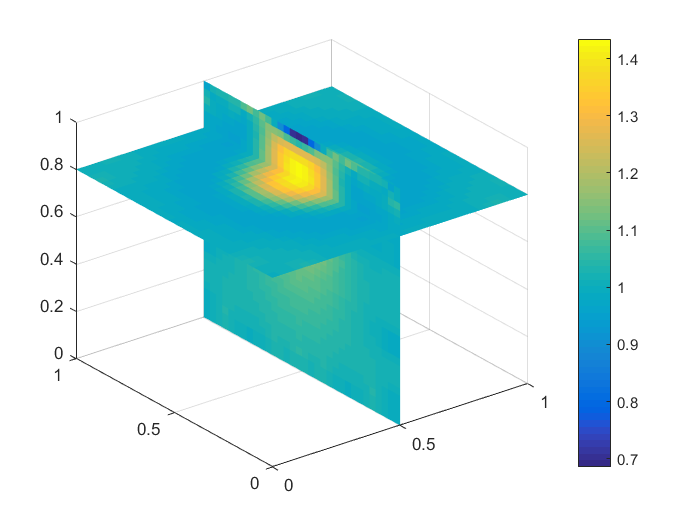} \\ 
(c) $N=1$ & (d) $N=1$ \\ 
\includegraphics[width=5cm]{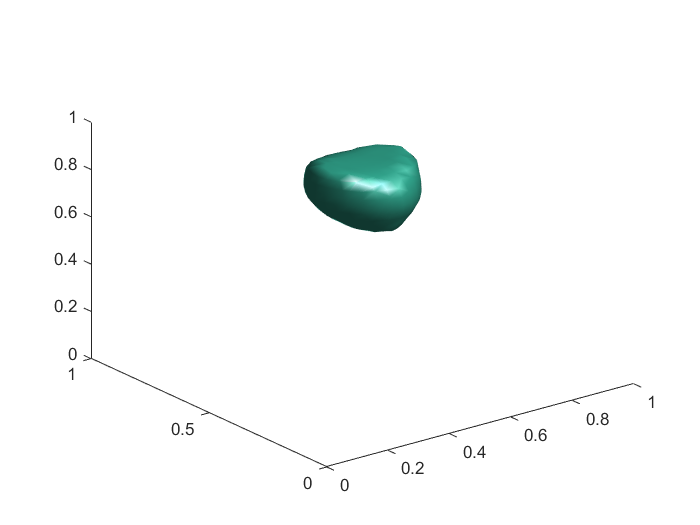} & %
\includegraphics[width=5cm]{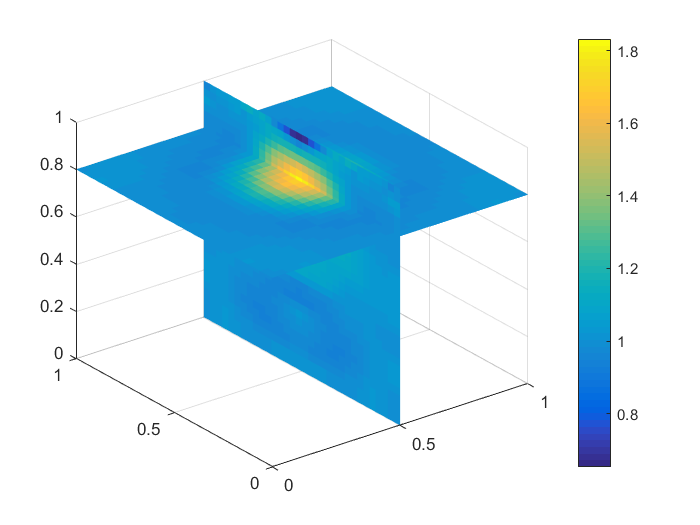} \\ 
(e) $N=3$ & (f) $N=3$%
\end{tabular}%
\end{center}
\caption{\emph{Results of Test 1.\ Imaging of one ball shaped inclusion with 
} $c=2$\emph{\ in it and $c=1$ outside. Hence, the inclusion/background
contrast is 2:1. We have stopped at the 3rd mesh refinement for all three
values of $N $. a) and b) Correct images. c) and d) Computed images for $N=1$%
. e) and f) Computed images for $N=3$. The maximal value of the computed
coefficient $c\left( \mathbf{x}\right) $ is approximately 1.8.}}
\label{example1}
\end{figure}

\textbf{Test 1}. First, we test the reconstruction by our method of a single
ball shaped inclusion depicted on Figure \ref{example1} a). $c=2$ inside of
this inclusion and $c=1$ outside. Hence, the inclusion/background contrast
is 2:1. We show the 3D image and slices for $N=1,3$, see Figures \ref%
{example1}.

\begin{figure}[tbp]
\begin{center}
\begin{tabular}{cc}
\includegraphics[width=5cm]{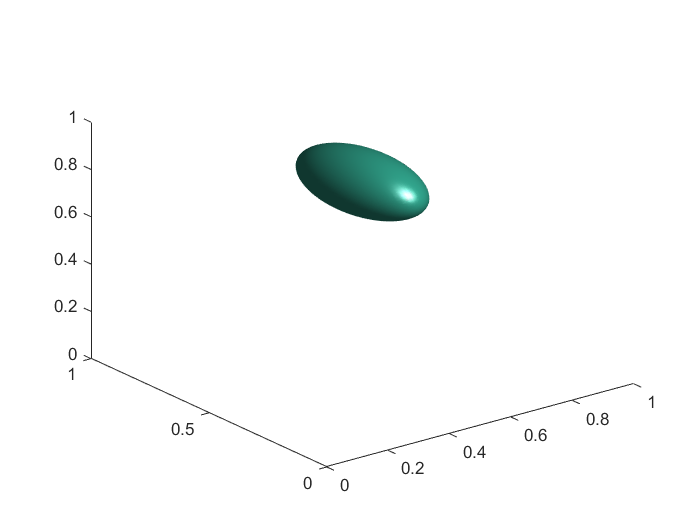} & %
\includegraphics[width=5cm]{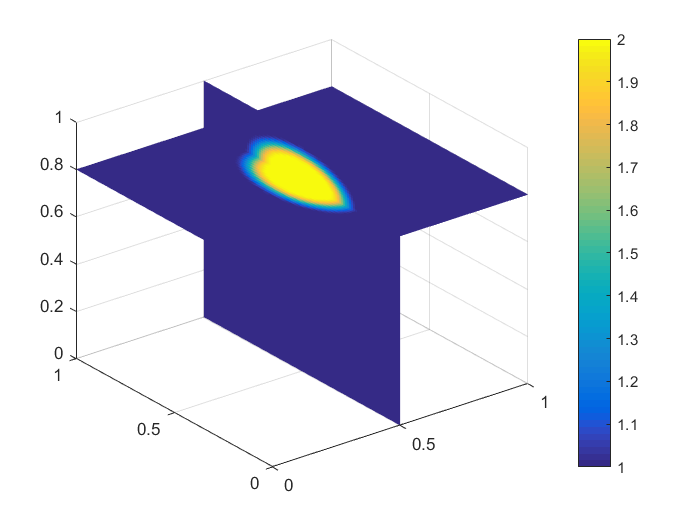} \\ 
(a)3D image of true $c$ & (b) Slice image of true $c$ \\ 
\includegraphics[width=5cm]{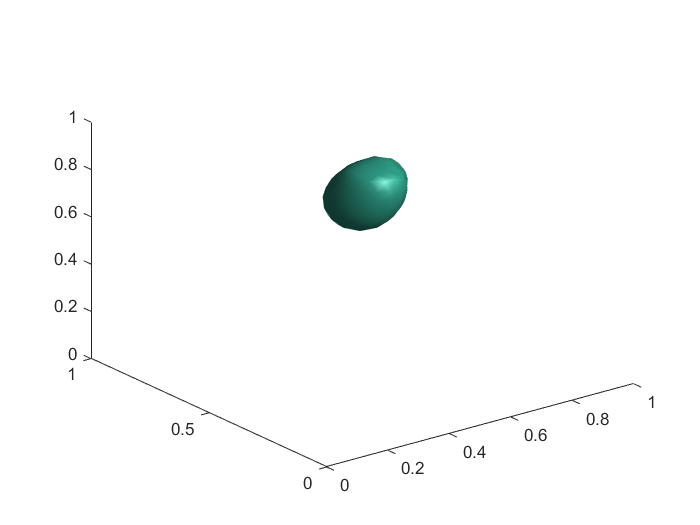} & %
\includegraphics[width=5cm]{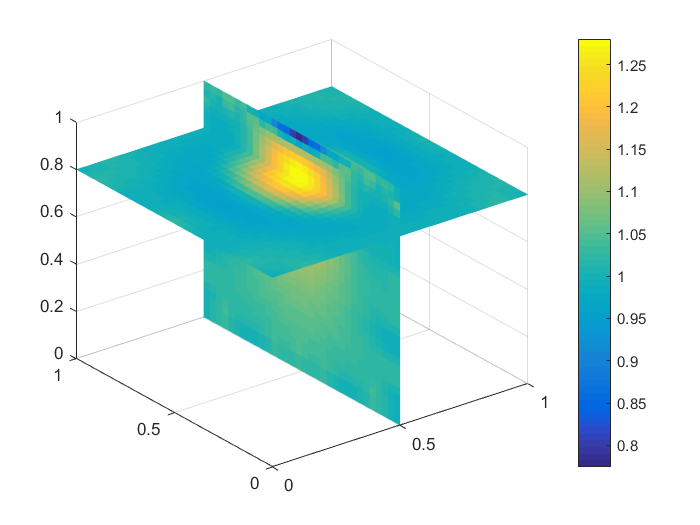} \\ 
(c) $N=1$ & (d) $N=1$ \\ 
\includegraphics[width=5cm]{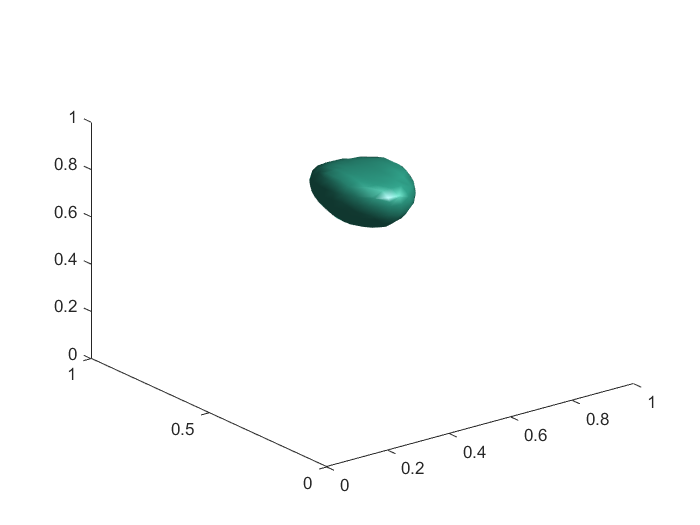} & %
\includegraphics[width=5cm]{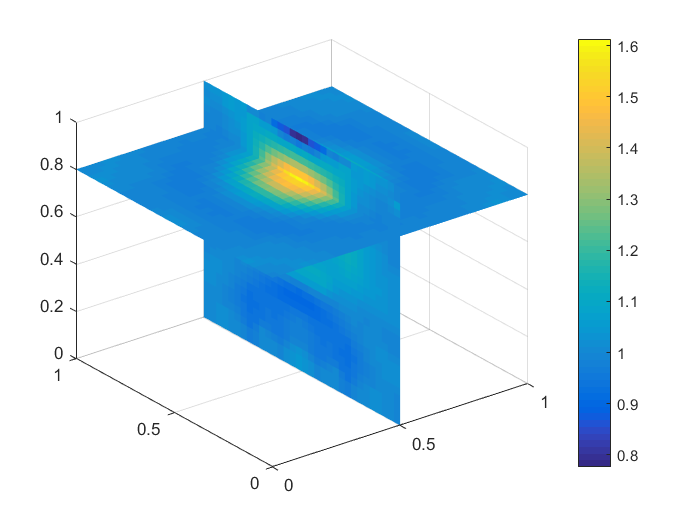} \\ 
(e) $N=3$ & (f) $N=3$%
\end{tabular}%
\end{center}
\caption{\emph{Results of Test 2.\ Imaging of one elliptically shaped
inclusion with } $c =2$\emph{\ in it and $c =1$ outside. Hence, the
inclusion/background contrast is 2:1. We have stopped at the 3rd mesh
refinement for all three values of $N$. a) and b) Correct images. c) and d)
Computed images for $N=1$. e) and f) Computed images for $N=3$. The maximal
value of the computed coefficient $c\left( \mathbf{x}\right) $ is
approximately 1.6.}}
\label{example2}
\end{figure}

\textbf{Test 2}. Second, we test the reconstruction by our method of a
single elliptically shaped inclusion depicted on Figure \ref{example2} a). $%
c=2$ inside of this inclusion and $c=1$ outside. Hence, the
inclusion/background contrast is 2:1. We show the 3D image and slices for $%
N=1,3$, see Figures \ref{example2}.

\begin{figure}[tbp]
\begin{center}
\begin{tabular}{cc}
\includegraphics[width=5cm]{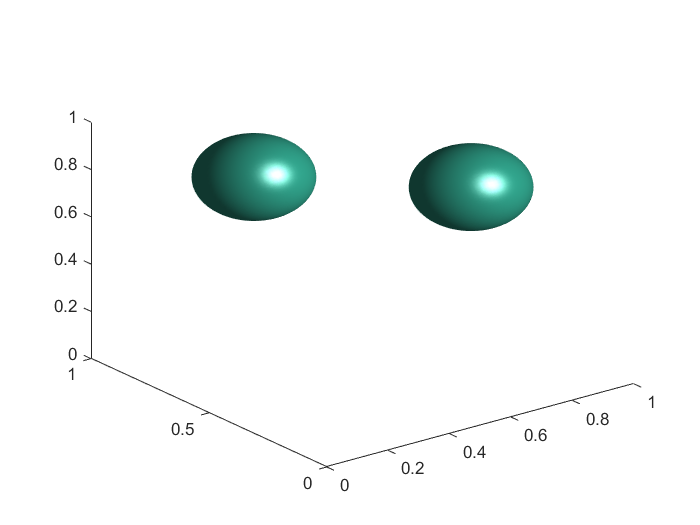} & %
\includegraphics[width=5cm]{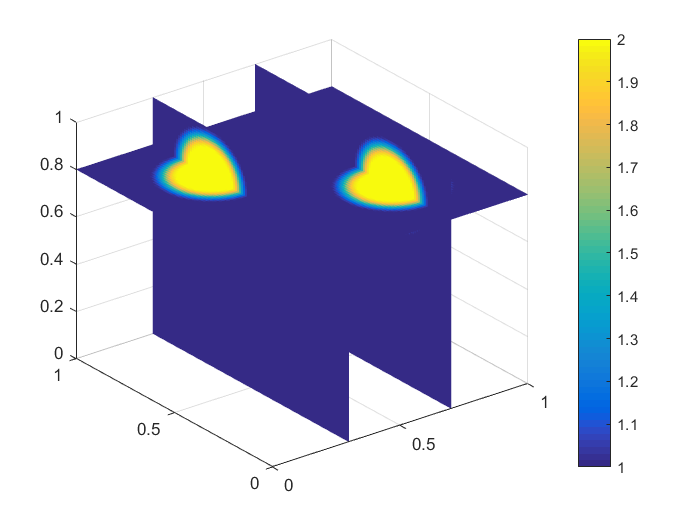} \\ 
(a)3D image of true $c$ & (b) Slice image of true $c$ \\ 
\includegraphics[width=5cm]{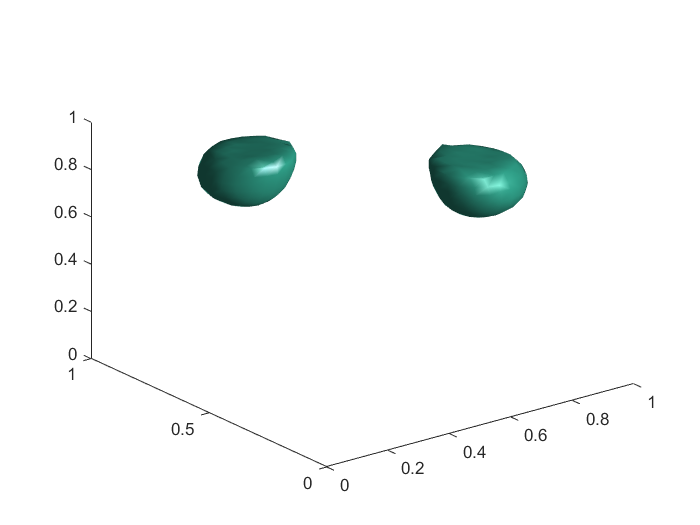} & %
\includegraphics[width=5cm]{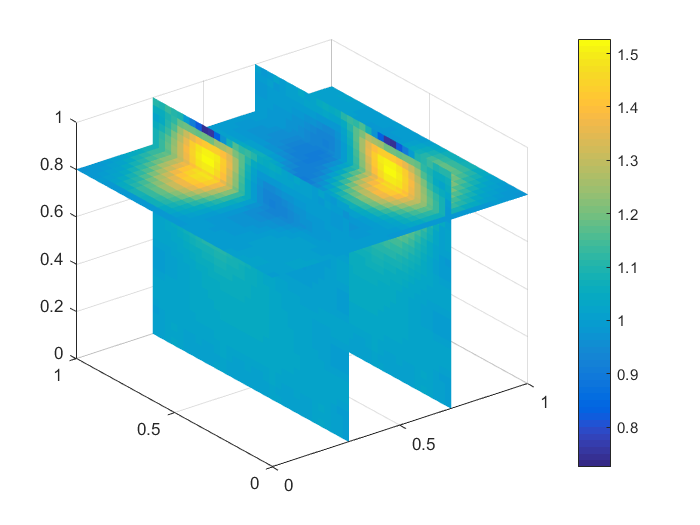} \\ 
(c) $N=1$ & (d) $N=1$ \\ 
\includegraphics[width=5cm]{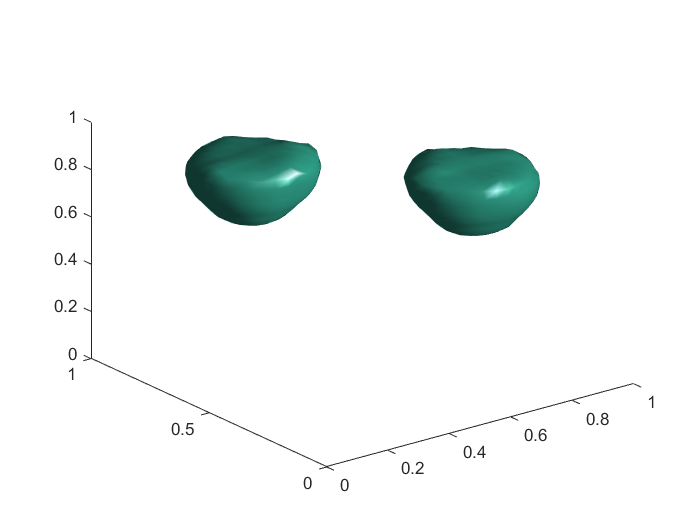} & %
\includegraphics[width=5cm]{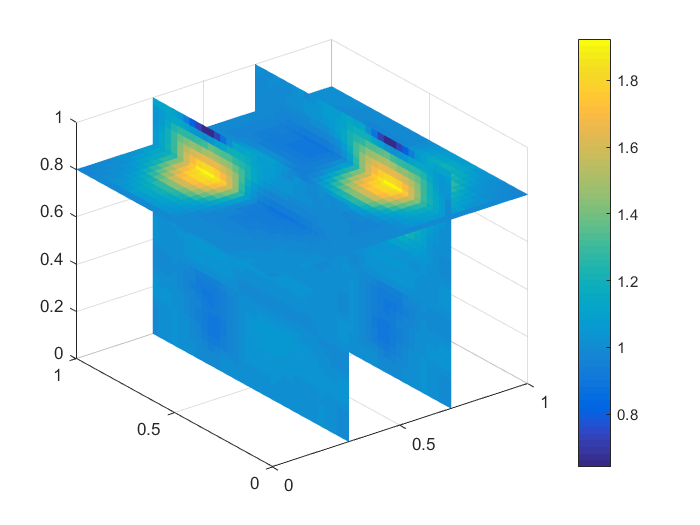} \\ 
(e) $N=3$ & (f) $N=3$%
\end{tabular}%
\end{center}
\caption{\emph{Results of Test 3.\ Imaging of two ball shaped inclusions
with } $c =2$\emph{\ in each of them and $c =1$ outside. We have stopped on
the 3rd mesh refinement for all three values of $N$. a) and b) Correct
images. c) and d) Computed images for $N=1$. e) and f) Computed images for $%
N=3$. In each imaged inclusion, the maximal value of the computed
coefficient $c\left( \mathbf{x}\right) $ is approximately 1.9.}}
\label{example3}
\end{figure}

\textbf{Test 3}. We now test the performance of our method for imaging of
two ball shaped inclusions depicted on Figure \ref{example3} a). $c=2$
inside of each inclusion and $c=1$ outside of these inclusions. Figures \ref%
{example3} display results.

\begin{figure}[tbp]
\begin{center}
\begin{tabular}{cc}
\includegraphics[width=5cm]{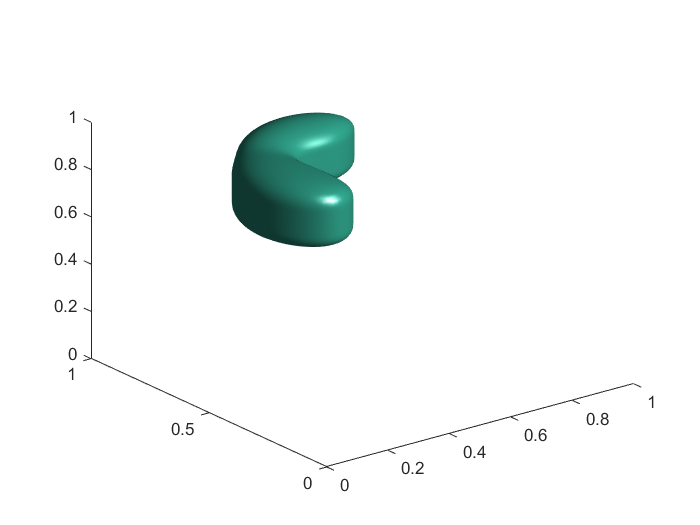} & %
\includegraphics[width=5cm]{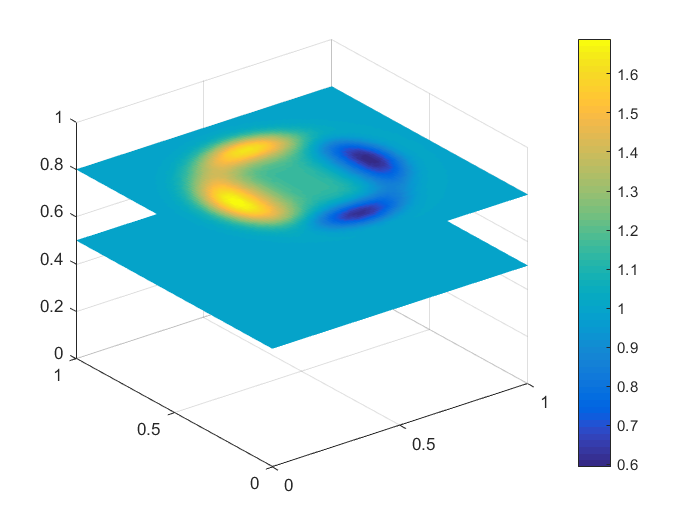} \\ 
(a) 3D image of true $c$ & (b) Slice image of true $c$ \\ 
\includegraphics[width=5cm]{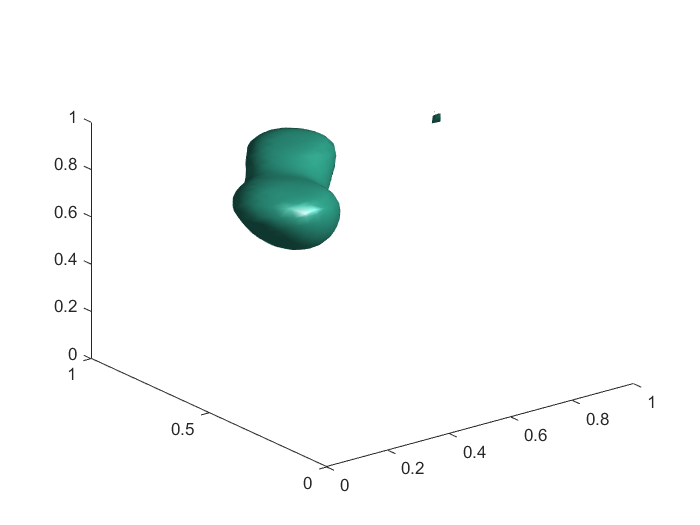} & %
\includegraphics[width=5cm]{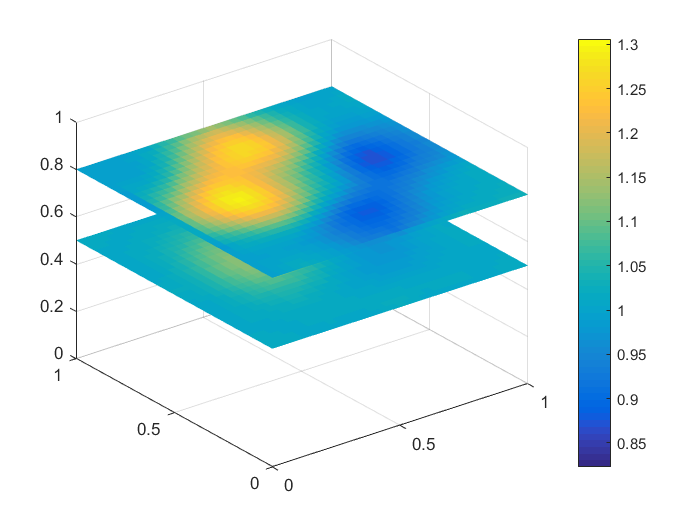} \\ 
(c) $N=1$ & (d) $N=1$ \\ 
\includegraphics[width=5cm]{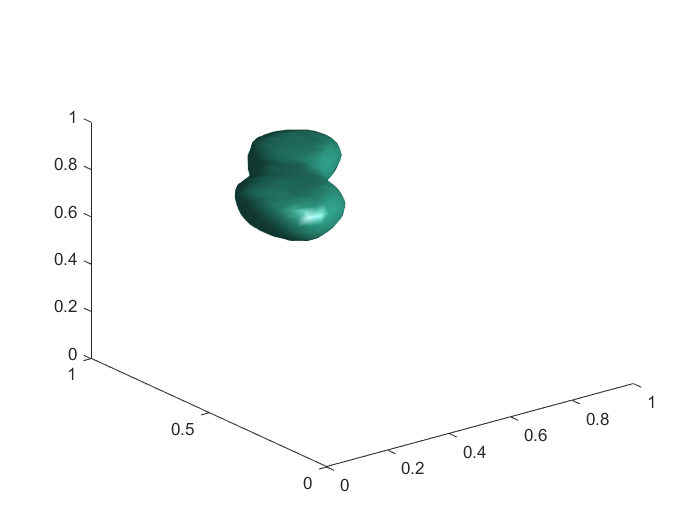} & %
\includegraphics[width=5cm]{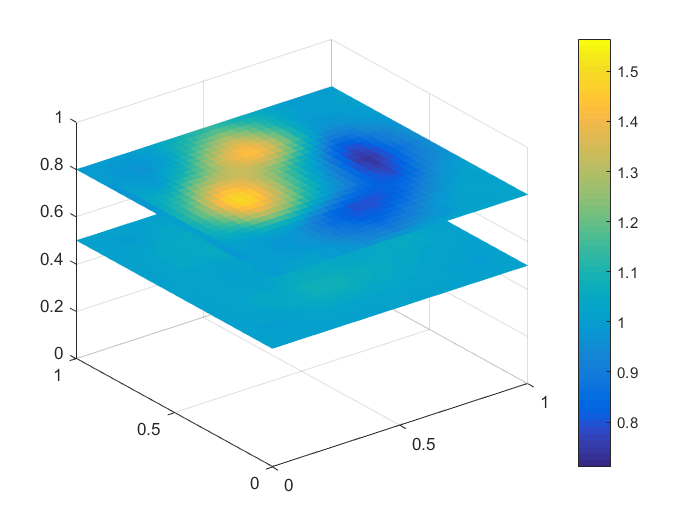} \\ 
(e) $N=3$ & (f) $N=3$%
\end{tabular}%
\end{center}
\caption{\emph{Results of Test 4. Imaging of a smoothly varying coefficient.
The function $c\left( \mathbf{x}\right) $ in the inclusion varies between
0.4 and 1.6. a) and b) Correct images. c) and d) Computed images for $N=1$.
e) and f) Computed images for $N=3$. The computed function $c\left( \mathbf{x%
}\right) $ in the inclusion varies approximately between 0.7 and 1.6.}}
\label{example4}
\end{figure}

\textbf{Test 4}. We now test our method for the case when the function $%
c\left( \mathbf{x}\right) $ is smoothly varying within an abnormality and
with a wide range of variations approximately between 0.6 and 1.7. The
results are shown in Figure \ref{example4}. Thus, our method can accurately
image not only \textquotedblleft sharp" inclusions as in Tests 1-3, but
abnormalities with smoothly varying functions $c\left( \mathbf{x}\right) $
in them as well.

\begin{figure}[tbp]
\begin{center}
\begin{tabular}{cc}
\includegraphics[width=5cm]{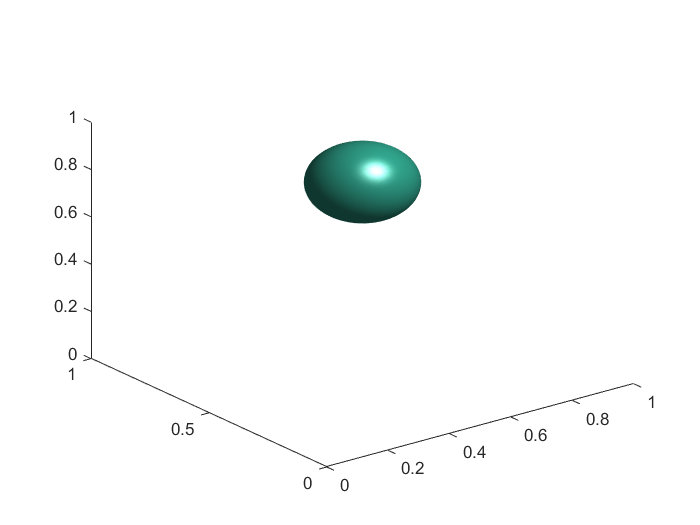} & %
\includegraphics[width=5cm]{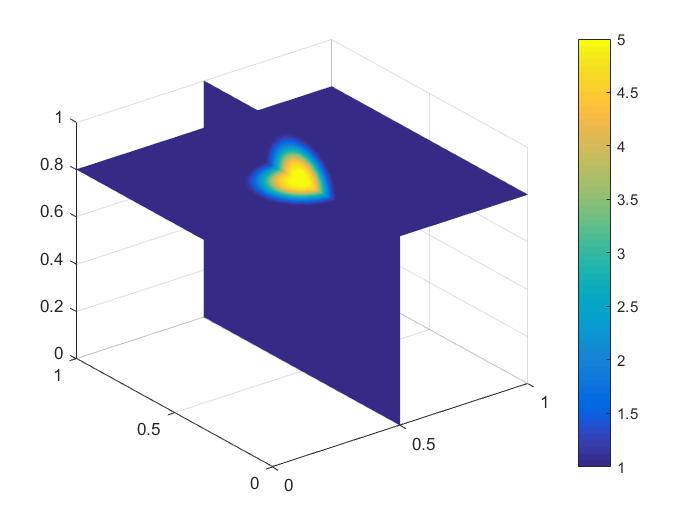} \\ 
(a) 3D image of true $c$ & (b) Slice image of true $c$ \\ 
\includegraphics[width=5cm]{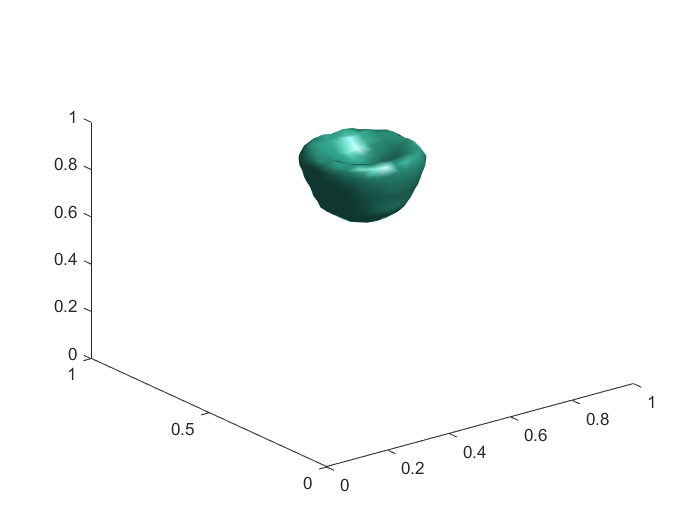} & %
\includegraphics[width=5cm]{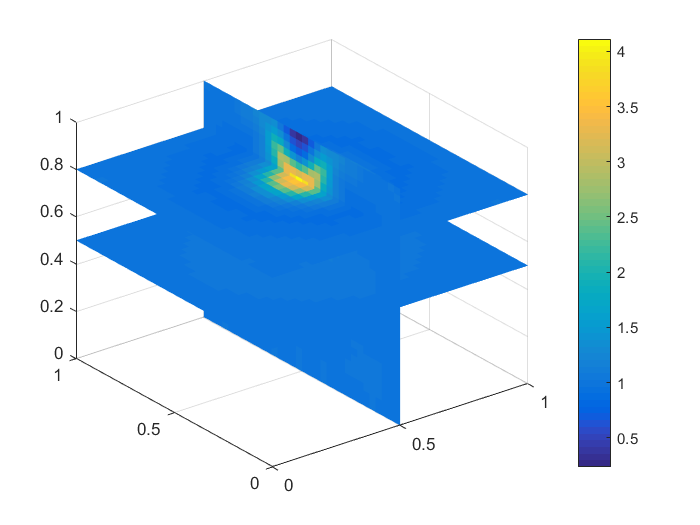} \\ 
(e) $N=3$ & (f) $N=3$%
\end{tabular}%
\end{center}
\caption{\emph{Results of Test 5.\ Imaging of one ball shaped inclusion with 
} $c =5$\emph{\ in it and }$c =1$ \emph{outside. Hence, the
inclusion/background contrast is 5:1. We have stopped at the 3rd mesh
refinement. a) and b) Correct images. c) and d) Computed images for $N=3$.
The maximal value of the computed coefficient $c\left( \mathbf{x}\right) $
is approximately 4.}}
\label{example5}
\end{figure}

\textbf{Test 5}. In this example, we test the reconstruction by our method
of a single ball shaped inclusion with a high inclusion/background contrast,
see Figure \ref{example5} a). $c=5$ inside of this inclusion and $c=1$
outside. Hence, the inclusion/background contrast is 5:1. See Figures \ref%
{example5} for results.

\textbf{Test 6}. In this example we test the stability of our algorithm with
respect to the random noise in the data. We test the stability for the case
of the function $c\left( \mathbf{x}\right) $ described in Test 4. The noise
is added for $\mathbf{x}\in \Gamma _{0}$ (see (\ref{6.0})) as: 
\begin{equation}
g_{0,\text{noise}}(\mathbf{x},t)=g_{0}(\mathbf{x},t)(1+\epsilon \xi _{t})%
\text{ and }g_{1,\text{noise}}(\mathbf{x},t)=g_{1}(\mathbf{x},t)(1+\epsilon
\xi _{t}),  \label{6.5}
\end{equation}%
where functions $g_{0}(\mathbf{x},t),g_{1}(\mathbf{x},t)$ are defined in (%
\ref{2.10}), $\epsilon $ is the noise level and $\xi _{t}$ is a random
variable depending only on the time $t$ and uniformly distributed on $[-1,1]$%
. We took $\epsilon =5\%$ which is 5\% noise.

\begin{figure}[tbp]
\begin{center}
\begin{tabular}{cc}
\includegraphics[width=5cm]{Figures/smmoth4_true_3d.png} & %
\includegraphics[width=5cm]{Figures/smooth4_true_slice.png} \\ 
(a) 3D image of true $c$ & (b) Slice image of true $c$ \\ 
\includegraphics[width=5cm]{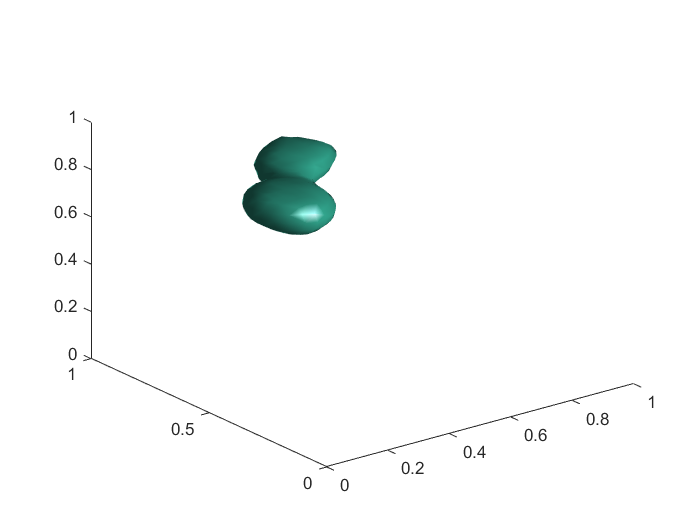} & %
\includegraphics[width=5cm]{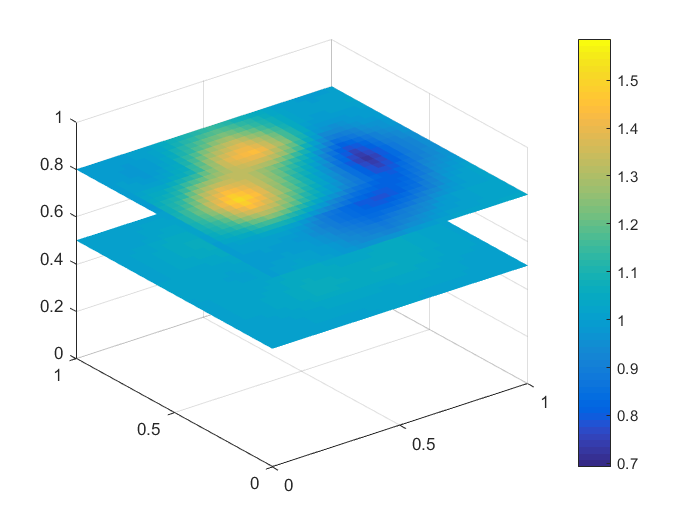} \\ 
(c) $N=3$ & (d) $N=3$%
\end{tabular}%
\end{center}
\caption{\emph{Results of Test 6. We test the reconstruction of the same
function $c\left( \mathbf{x}\right) $ as in Test 4 (Figures 5) but with the
noise in the data. The level of noise in \eqref{6.5} is $\protect\epsilon%
=5\% $. We have stopped at the 3rd mesh refinement for $N=3$. a) and b)
Correct images. c) and d) Computed images for $N=3$. The computed function $%
c\left( \mathbf{x}\right) $ in the inclusion varies between 0.7 and 1.6.}}
\label{example6}
\end{figure}

\end{document}